\documentclass[manuscript]{aastex}
\usepackage{amsmath}
\usepackage{xspace}
\usepackage{graphicx}
\usepackage{CJK}
\newcommand\luv{$L_{\rm uv}$}
\newcommand\lbol{$L_{\rm bol}$}
\newcommand\J{\textit{J}}
\newcommand\Herschel{\textit{Herschel}}
\shorttitle{UV-heated outflow walls}
\shortauthors{Lee et al.}

\begin{document}

\title{A PDR model for the FIR mid-$J$ CO ladder with universal rotational temperature in star forming region}

\author{Seokho Lee }
\affil{Astronomy Program, Department of Physics and Astronomy, Seoul National University, 1 Gwanak-ro, Gwanak-gu, Seoul 151-742, Korea}
\email{shlee@astro.snu.ac.kr}

\author{Jeong-Eun Lee}
\affil{Department of Astronomy and Space Science, Kyung Hee University, Yongin-shi, Kyungki-do 449-701, Korea}

\author{Edwin A. Bergin}
\affil{Department of Astronomy, University of Michigan, 830 Dennison Building, 500 Church Street, Ann Arbor, MI 48109, USA}

\and
\author{Yong-Sun Park}
\affil{Astronomy Program, Department of Physics and Astronomy, Seoul National University, 1 Gwanak-ro, Gwanak-gu, Seoul 151-742, Korea}

\begin{abstract}
 A photon dominated region (PDR) is one of the leading candidate mechanisms for the origin of the warm CO gas with near universal $\sim$300~K rotational temperature inferred from the CO emission detected towards embedded protostars by \Herschel/PACS. We have developed a PDR model in  general coordinates, where we can use the most adequate coordinate system for an embedded protostar having outflow cavity walls, to solve  chemistry and gas energetics self-consistently for given UV radiation fields with different spectral shapes. Simple 1D tests and applications show that FIR mid-\J\,($14\leq J\leq 24$) CO lines are emitted from near the surface of a dense region exposed to high UV fluxes. We apply our model to HH46 and find the UV-heated outflow cavity wall can reproduce the mid-\J\, CO transitions observed by \Herschel/PACS. A model with UV radiation corresponding to a blackbody of 10,000~K results in the rotational temperature lower than 300~K, while models with the Draine interstellar radiation field and the 15,000 K blackbody radiation field predict the rotational temperature similar to the observed one.
\end{abstract}
%
%
\section{Introduction}\label{sec:intro}
Many energetic phenomena, such as high energy photons produced from accretion onto a protostar and jets ejected from the star-disk boundary region, affect the physical and chemical structure of the disk and envelope simultaneously. This material is heated to a temperature from $\sim$100 to $\sim$1,000~K, where many key gas coolants are excited to emit in the far-infrared (FIR); in this respect CO is one of the most important coolants.

 Low-mass embedded protostars were observed with the Long Wavelength Spectrometer \citep[LWS, ][]{clegg96} aboard the Infrared Space Observatory \citep[ISO, e.g.,][]{Benedettini2003,vandishoeck04}. The CO rotational temperature $T_{\rm rot}$  obtained by fitting the CO excitation diagrams (up to \J~=~19--18, $E_{\rm up}$~=~1,050~K) were a few hundred to $\sim$1,000~K.
 Because of the low spatial resolution of ISO, however, the heating mechanism of CO gas (high energy photons or shocks) was not well constrained. 

 More recently, the observations of embedded low mass protostars with the Photodetector Array Camera and Spectrometer \citep[PACS; ][]{poglitsch10} aboard the \textit{Herschel} Space Observatory (\Herschel) revealed two temperature (warm and hot) CO gas components \citep{Manoj2013,karska13,Green2013}, which may be attributed to photon dominated region (PDR) and shock, respectively. \citet{Visser12} showed that the warm component of CO gas with $T_{\rm rot} \sim$~300~K can be produced by the PDR along the outflow cavity walls combined with a C-shock by modeling the CO fluxes detected with PACS. \citet{Visser12} also showed that the contribution of PDR to the CO emission increases with evolution. 

Many theoretical PDR models have been developed for three decades \citep[e.g.,][hereafter R07]{Rollig07}. Some codes deal with the detailed microphysics needed to model both chemistry and thermal balance \citep[e.g.,][]{LePe06, LePe09}, while others use approximate formulae or a reduced chemical network \citep[e.g.,][]{Rollig06, bruderer2009, Woit09}. The results of these models, therefore, spread out up to 1 dex in the predicted thermal structure in the far-ultra violet (FUV) irradiated gas.
 
 Most PDR models have concentrated on bright dense quiescent molecular gas exposed to radiation from O stars. However, FUV observations and theoretical models of classical T-Tauri stars show that these sources emit FUV radiation approximated by a 10$^4$~K blackbody radiation (hereafter BB1.0) produced mostly by accretion \citep[e.g., ][]{Gullbring1998, Calvet1998, Johns-Krull2000,yang12}. This FUV spectrum with a lower effective temperature than those of O stars affects the composition and structure of PDRs \citep{Spaans94} because the reduction in the FUV radiation at the shortest wavelengths (912 -- 1,100\AA) reduces the efficiency of the photoelectric heating on polycyclic aromatic hydrocarbons (PAHs) and small dust grains \citep{Spaans94}, and also reduces the photodissociation rates of H$_2$ and CO \citep{vandishoeck06}.

 The PDR model for the embedded protostar with outflow cavity walls needs to deal with an at least two dimensional system and to cover a high dynamic range of physical parameters, at  radii from $\sim$10 AU to $\sim$10$^4$ AU. Recently, some PDR models have started to consider the requisite two-dimensional geometries \citep[e.g., ][]{vanzadelhoff2003,bruderer2009,Woit09}. These 2D PDR models use the cylindrical coordinate system concentrating on the protoplanetary disk. However, the cylindrical coordinate system needs a large number of grids, and thus, increases computational time to model the outflow cavity walls of embedded protostars with a reasonable spatial resolution. For example, \citet{bruderer2009} modeled the UV heated outflow cavity walls with $\sim$10$^5$ grid cells.
 
In this paper, we apply a new PDR code to the two-dimensional density structure of embedded outflow sources combined with a 15,000~K blackbody FUV radiation field (hereafter BB1.5), fitted to the observed UV spectrum of TW Hya \citep[e.g., ][]{Herczeg2002, yang12}, as well as BB1.0 and the Draine field. In Section~\ref{sec:model}, we describe in detail the ray tracing in the general grid, the FUV radiative transfer, chemistry, and gas energetics adopted in our new PDR model. In Section~\ref{sec:benchmark}, we test the newly developed PDR code with the benchmark models described by R07 and compare with other published codes. We present the FIR CO lines produced by the PDR model in Section \ref{sec:co1d} and apply our 2D PDR code to the CO ladder observations of HH46 in Section~\ref{sec:application}. Finally, we summarize our work in Section \ref{sec:summary}.

%
%

\section{Model}\label{sec:model}
\begin{figure*}
    \includegraphics[width=0.5 \textwidth]{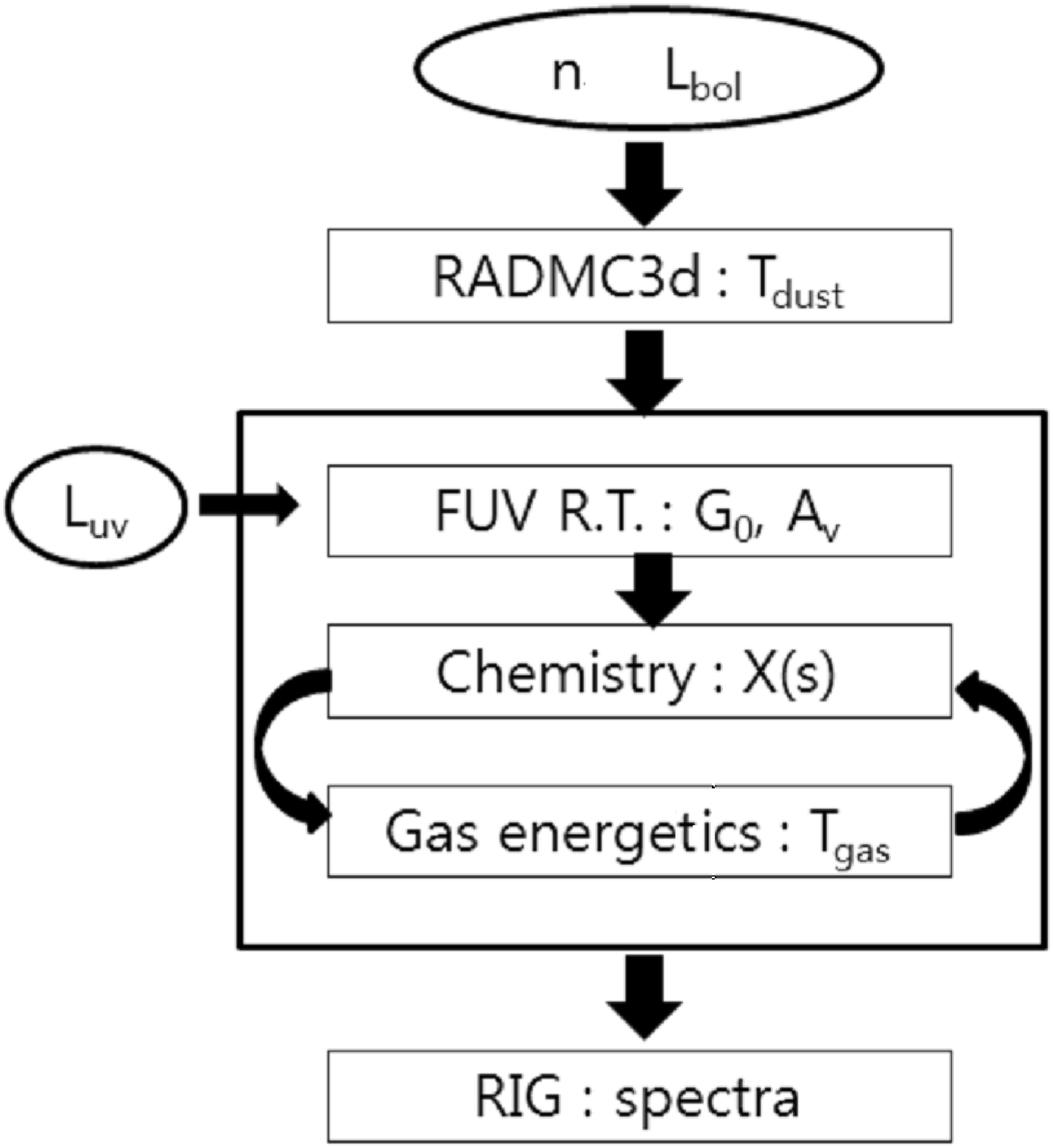}
    \caption{ Model procedure. Free parameters are the density distribution $n$, bolometric luminosity \lbol, and UV luminosity \luv. We find the converged solution of the chemistry and the gas energetics in the PDR model and synthesized the line spectra by using non-LTE line Radiative transfer code In General grid (RIG).}\label{fig:flowchart}
\end{figure*}

Our newly developed PDR code solves the FUV radiative transfer, chemistry, and gas energetics self-consistently. The procedures of our model are summarized in Fig. \ref {fig:flowchart}. First, for a given density structure, the dust temperature $T_{\rm dust}$ is calculated with the dust continuum radiative code RADMC-3D\footnote{http://www.ita.uni-heidelberg.de/$^\sim$dullemond/software/radmc-3d/}. Next, in the PDR model, we calculate the FUV radiative transfer to get unattenuated FUV strength $G_0$ and average visual extinction $\left<A_{\rm V}\right>$, and then solve chemistry and gas energetics iteratively. Finally, we synthesize molecular lines with a non-local thermal equilibrium (LTE) line radiative transfer code to compare with observations. Each part of our PDR model is described in detail below.

\subsection{Ray tracing in general grids}\label{sec:raytrace}
 We adopt a grid-based Monte Carlo method, which is a very flexible method to solve the radiative transfer and can take the anisotropic scattering from dust grains in the FUV radiative transfer into account easily. Some PDR codes considered only isotropic scattering \citep[e.g., prodimo,][]{Woit09} or  the extinction without considering the scattering \citep[3D-PDR,][]{bisbas12} to reduce the computational time. However, \citet{Rollig13} showed that isotropic and anisotropic scattering can produce flux differences of about 20~\% near the surface and a factor of two in the deeper region ($A_{\rm V}$ $\sim$ 5). 

In the grid-based radiative transfer, we need to know only the distance to the nearest surface of a grid for a given photon propagation direction. When a photon propagates as much as $ds$, a trajectory of the photon is described in the Cartesian coordinate as 
\begin{eqnarray} 
\label{eq:xpos}
\overrightarrow{X} &=& \overrightarrow{X_0} + \widehat{X} \cdot ds \\ \nonumber
     (x,y,z)       &=& (x_0,y_0,z_0) + (\hat{x},\hat{y},\hat{z}) \cdot ds,
\end{eqnarray}
where $\overrightarrow{X_0}$ is the current position, $\overrightarrow{X}$ is the next position, and $\widehat{X}$ is the direction vector.  Because the surface of the grid can be described by a simple equation with $x$, $y$, and $z$ in the Cartesian coordinates, we can find $ds$ by solving the equation of the photon trajectory intersecting the surface of the grid in any coordinate system.

 For example, the boundary between the outflow cavity and the envelope can be described by 
\begin{eqnarray}\label{eq:bound}
 z & = & \delta_0 \times (x^2+y^2) \\ \nonumber
   & = & \left( \frac{1}{10^4\,\rm{AU} \tan^2(\alpha/2)}\right) \times (x^2 + y^2),
 \end{eqnarray}
 where $z$ is the outflow axis and $\alpha$ is the full opening angle at $z$~=~$10^4$~AU \citep{bruderer2009}. As the boundary parameter $\delta_0$ describes a circular paraboloid, the circular paraboloid with $\delta \equiv z / (x^2 +y^2)$ can be used as a new coordinate instead of a circular conical surface $\theta$ in the spherical coordinates. In this ($r$,~$\delta$) coordinates, using Eq. \ref{eq:xpos} and the definition of $\delta$, we find the quadratic equation of the photon trajectory intersecting the $\delta$ surface as
\begin{equation}
\label{eq:quad}
A\cdot ds^2 + B\cdot ds + C =0,
\end{equation}
where
\begin{eqnarray}
A & = & \delta \left(\hat{x}^2 + \hat{y}^2 \right) \nonumber\\ 
B & = & 2 \delta \left(\hat{x} x_0 + \hat{y} y_0 \right) - \hat{z}\nonumber\\ 
C & = & \delta \left(x_0^2 + y_0^2 \right) -z_0^2. \nonumber
\end{eqnarray}

Therefore, in order to minimize the computational time, we can choose a coordinate system optimized to a given physical model, which  can provide an enough spatial resolution with a relatively small number of grids (see Sec.~\ref{sec:application} for more detail).

\subsection{FUV radiative transfer}\label{sec:fuvrt}
 The FUV radiative transfer is calculated by the method of \citet{vanzadelhoff2003} and \citet{bruderer2009}. We calculate the FUV radiative transfer at only one representative wavelength where photon energy is 9.8~eV (the middle of the 6~-~13.6~eV FUV band) and then measure the FUV strength (the unattenuated FUV strength $G_0$ and the attenuated FUV strength $G_{\rm dust}$) in units of the Habing field \citep[ISRF, $1.6 \times 10^{-3}\, {\rm erg\, s^{-1}\, cm^{-2}}$,][]{Habing1968}. Therefore, BB1.0 and BB1.5 are normalized to have the same integrated intensity from 912--2050 \AA\, as ISRF, and the Draine field \citep[$\chi$,][]{Draine1978} is given by $\chi\, =\, G_{\rm dust} /\, 1.71$. We adopt dust properties for the average Milky Way dust in molecular clouds with $R_{\rm V}$~=~5.5 and C/H~=~48~ppm in PAHs \citep{Draine03} for this calculation.
 
  To derive the unattenuated FUV strength $G_0$ and the attenuated FUV strength $G_{\rm dust}$ in the 2D space, we solve the FUV radiative transfer with the dust scattering using Henyey-Greenstein phase function,
\begin{equation} 
 P(\cos \phi,g_{\lambda})=\frac{1-g^{2}_{\lambda}}{4 \pi[1+g_\lambda^{2}-2g_{\lambda}\cos \phi]^{3/2}}
\label{eq: henyey}
\end{equation}
with the mean scattering angle $g_{\lambda}~=~<\rm{cos}\phi>~=~0.767$. The scattering optical depth is first calculated using random number $\zeta$ between 0~and~1 as
\begin{equation}
\tau_{\rm scat} = -\ln (1-\zeta),
\end{equation}
which can be converted to an absorption optical depth, 
\begin{equation}
\tau_{\rm abs} = \tau_{\rm scat} \times\omega /(1 - \omega)
\end{equation}
with the dust grain albedo $\omega$~=~0.387.

 Each model photon has the initial intensity $I(0)$ given by
\begin{equation}
I(0) = \frac{F \cdot S}{N_{\rm phot}}
\end{equation} 
where $F$ is the flux entering the system, $S$ is the total surface that the photon passes through, and $N_{\rm phot}$ is the number of model photons. The model photon propagates until it reaches the optical depth ($\tau_{\rm scat}$) at which it scatters, and its intensity drops according to
\begin{equation}
I_i(s+\Delta s) = I_i(s) \exp(-\Delta \tau_{\rm abs}) 
\end{equation}
\begin{equation}
\Delta \tau_{\rm abs} =(1 - \omega) \, C_{\rm ext}\, n\,\Delta s,
\end{equation}
where $C_{\rm ext}$ is an extinction cross section of 1.075$\times$10$^{-21}$~cm$^{2}$ per H nucleus,  $n~(=~n_{\rm H}~+~2n_{\rm H_2})$ is the total hydrogen number density, and $\Delta s$ is the path length traveled within a grid cell. 
Therefore, the dust attenuated FUV strength $G_{\rm dust}$ in a grid cell with the volume V is
 \begin{equation}
 G_{\rm dust} =\frac{1}{1.6\times 10^{-3}\, {\rm erg}\, {\rm cm}^{-2}\, {\rm s}^{-1}}\frac{1}{V}\sum I_i \Delta s \frac { ( 1 -\exp(-\Delta \tau_{\rm abs}))}{\Delta \tau_{\rm abs}}
 \end{equation}
 where the sum is taken over all photon packages passing the grid cell. The unattenuated 
 FUV strength $G_{\rm 0}$ in the grid cell with the volume $V$ is 
 \begin{equation}
  G_{\rm 0} =\frac{1}{1.6\times 10^{-3}\, {\rm erg}\, {\rm cm}^{-2}\, {\rm s}^{-1}}\frac{1}{V}\sum I_0 \Delta s,
 \end{equation}
and the average visual extinction $\left<A_{\rm V}\right>$ is 
 \begin{equation} 
 \left<A_{\rm V}\right> = - \ln \left(\frac{G_{\rm  dust}}{G_0}\right)\frac{1}{2.5\mathrm{log(e)}}\frac{1}{k_{\rm UV/V}},
 \label{eq:Av}
 \end{equation}
where the conversion factor of $k_{\rm UV / V} \left(= A_{\rm UV}/A_V\right)$ is 1.6.
 As each photon passing the grid cell comes through a different column density, $\left<A_{\rm V}\right>$ is the mean over all photons. $A_{\rm V}$ is the visual extinction in 1D model and related directly to the column density while $\left<A_{\rm V}\right>$ is calculated with Eq.~\ref{eq:Av} in 2D model and averaged over all photons.
  We note that  $G_0$ is calculated by neglecting absorption by dust grains, i.e., scattering by grains is still considered. Because otherwise, $G_{\rm dust}$ is larger than $G_0$ in some cases, resulting in a minus value of $\left<A_{\rm V}\right>$. To prevent this effect, we define $G_0$ as a FUV strength in the absence of only absorption by grains, as following \citet{bruderer2009}.

 For the convergence, the number of model photons are doubled until the difference in  $G_{\rm dust}$ in two consecutive steps, $\left|G_{\rm dust}({\rm previous})- G_{\rm dust}({\rm current})\right| / G_{\rm dust}({\rm current})$, is smaller than 5~\% for all grid cells where $G_{\rm dust}$ is larger than $10^{-4}$ ISRF. 

%
%

\subsection{Chemistry} \label{sec:chem}
 For chemistry, we have modified the Heidelberg ``ALCHEMIC'' code \citep{semenov2010}.
 The basic equations for species in the gas phase ($n(i)$) are described as
\begin{eqnarray}
  \frac{d n(i)}{dt} &=& -n(i) \sum{n(j) k_{\rm g2}^{i,j}} + \sum{n(j)n(k)k_{\rm g2}^{j,k}} \\ \nonumber
                    & &  -n(i) \sum{k_{\rm g1}^{i}} + \sum{n(j)k_{\rm g1}^{j}}\\ \nonumber
                    & &  -n(i)k_{\rm ad} + n_s(i) k_{\rm des}. 
\end{eqnarray}
In the equation, the first term indicates the destruction process of the given species $n(i)$ by reacting with another species, and the second term describes the formation process by the reactions of two other species. The third and fourth terms represent destruction and formation of the species, respectively, through photodissociation/ionization and cosmic ray ionization. The last two terms describe the adsorption and desorption of the species onto and out of grain surfaces, respectively.
   
The gas-phase chemical reaction network is based on UMIST2006 database \citep{UMIST2007} modified by \citet{Bruderer09a}. The two body reaction rate is expressed as 
  \begin{equation}
  k_{\rm g2}  = \alpha_{\rm g2} \times \left(\frac{T_{\rm gas}}{300 {\rm K}}\right)^{\beta_{\rm g2}} \exp\left(-\gamma_{\rm g2}/T_{\rm gas}\right) \quad \mathrm{cm}^{3}\, \mathrm{s}^{-1}
  \end{equation}
  where $\alpha_{\rm g2}$, $\beta_{\rm g2}$, and $\gamma_{\rm g2}$ are coefficients that depend on reaction types.
   
The FUV photoreaction rate is described as
\begin{equation}
  k_{\rm ph} = \chi \alpha_{\rm ph} \exp \left(-\gamma_{\rm ph} A_{\rm V} \right) \quad \mathrm{s}^{-1}
\end{equation} 
  where $\chi$ is the FUV strength in Draine field.  
  Unshielded rates ($\alpha_{\rm ph}$) are calculated with the cross sections given by \citet{vandishoeck06}, and dust attenuation factors, $\gamma_{\rm ph}$ for $R_{\rm V}$~=~5.5 grain are adjusted by the method of \citet{Rollig13}. 
  Unshielded photo-dissociation rates of H$_2$ and CO in BB1.0 are $3.16 \times 10^{-12} $~s$^{-1}$ and  $1.90 \times 10^{-11}$~s$^{-1}$, respectively, which are lower than the rates in the Draine field by an order of magnitude because the intensity between 912--1100 \AA\,  in BB1.0 is lower than that of the Draine field by an order of magnitude \citep{vandishoeck06}.
 
 Self-shielding of H$_2$ and CO cause the rapid decrease of their photodissociation. The approximate formula for the H$_2$ self-shielding is given by: 
\begin{eqnarray}\label{eq:H2-shield}
\beta _{\rm  H2} & =& \frac{0.965}{(1 + x_{\rm H2}/b_5)^2}+\frac{0.035}{(1+x_{\rm H2})^{0.5}}\\ \nonumber
                  &&\times {\rm exp}[-8.5 \times 10^{-4} (1+x_{\rm H2})^{0.5} ],
\end{eqnarray}
where $x_{\rm  H2} \equiv N_{\rm  H2}/5\times 10^{14} {\rm cm}^{-2}$ and $b_5 \equiv b/10^5 {\rm cm\, s^{-1}}$ \citep{Draine96}. Here, $N_{\rm H2}$ is the H$_2$ column density, and $b$ is the Doppler broadening parameter ($b \equiv \mathrm{FWHM}/\sqrt{4{\rm ln} 2}$), which is assumed as 1.1~km s$^{-1}$.  For the CO self-shielding effect, we interpolate the values on Table 6 ($b$(CO) = 0.3 km s$^{-1}$, $T_{\rm  ex}$(CO) = 50 K) in \citet{visser09}. Neutral carbon is also shielded by H$_2$ in addition to the self-shielding, which is taken into account by a simple factor \citep{Kamp2000,Woit09}:
\begin{equation}
\label{eq:beta_c}
\beta_{\rm C} = \exp\left( -\sigma_{\rm C}^{\rm bf}N_{\rm C} - 0.9 T_{\rm gas}^{0.27} \left(\frac{N_{\rm H_2}}{10^{22} \mathrm{cm}^{-2}}\right)^{0.45}\right)
\end{equation}
with the neutral carbon column density, $N_{\rm C}$ and the FUV-averaged cross section of the neutral carbon, $\sigma_{\rm C}^{\rm bf} = 1.1 \times 10^{-17} \mathrm{cm}^{2}$.

In a deeper region of PDR, where most FUV photons are shielded, the cosmic ray affects chemistry significantly. The cosmic ray ionization reaction rate ($k_{\rm CR}$) is given by,
\begin{equation}
 k_{\rm CR} = \frac {\zeta_{\rm CR}}{1.36 \times 10^{-17} \mathrm{s}^{-1}} \alpha_{\rm CR} \quad \mathrm{s}^{-1},
\end{equation}
and the cosmic ray induced photoreaction rate ($k_{\rm CRP}$)  is given by,
\begin{equation}
k_{\rm CRP} =  \frac {\zeta_{\rm CR}}{1.36 \times 10^{-17} \mathrm{s}^{-1}} \alpha_{\rm CR} \left(\frac{T_{\rm gas}}{300 {\rm K}}\right)^{\beta_{\rm CRP}} \frac{\gamma_{\rm CRP}}{1-\omega} \quad  \mathrm{s}^{-1}
\end{equation}
where $\zeta_{\rm CR}$ is the cosmic ray ionization rate of ${\rm H}_2$, $\alpha_{\rm CR}$ is the cosmic ray ionization rate of the given species, $\gamma_{\rm CRR}$ is the efficiency of cosmic ray ionization event, and $\omega$ is the grain albedo assumed to be 0.5. 
The cosmic ray is attenuated by grains with the attenuation column density of 96 $\rm{g\,cm}^{-2}$ \citep{Umebayashi1981,fogel11}. Though $\zeta_{\rm CR}$ is recently estimated in diffuse clouds as $3.5 \times 10^{-16} \mathrm{s}^{-1}$ \citep{Indriolo2012}, $\zeta_{\rm CR}$ in dense molecular cores is a few times $10^{-17} \mathrm{s}^{-1}$ \citep{van der Tak2000,Hezareh2008}. As protostars are generally embedded in dense molecular cores, we adopt $\zeta_{\rm CR} = 5 \times 10^{-17}$ s$^{-1}$ \citep{dalgarno06}.

 Neutral gas species freeze out to grain and evaporate from  the grain. We assume the grain size ($a_{\rm gr}$) of  0.1 $\mu$m and adopt binding energies and photo-desorption yields from \citet{fogel11}. The binding energies and photon yields for some important species are listed in Table~\ref{tab:BePy_grain}. 
  
The adsorption of a species onto grains is given by
\begin{equation}
\label{eqn-adsorption}
k_{\rm ad}\, =\, \sigma_{\rm gr}\, \sqrt{\frac{8.0 k_{\rm B} T_{\rm  gas}}{\pi \mu m_H}}\, S_{\rm sp}\, n_{\rm gr} \quad \mathrm{s}^{-1},
\end{equation}
where $\sigma_{\rm gr}$ is the cross section of the dust grain ($\pi a_{\rm gr}^2=3.14 \times 10^{-10} $~cm$^{2}$),  $k_{\rm B}$ is Boltzmann's constant, $T_{\rm  gas}$ is the gas temperature, $\mu$ is the molecular weight of the species, $m_{\rm  H}$ is the mass of atomic hydrogen, $n_{\rm gr}$ is the number density of grains, and $S_{\rm sp}$ is the sticking coefficient, assumed to be unity for all species.

The thermal desorption of a species from grains is calculated using the Polyani-Wigner relation:
\begin{equation}
k_{\rm td} = \nu_{\rm s}\, e^{-E_{\rm b}/T_{\rm  dust}} \quad \mathrm{s}^{-1}
\end{equation}
where $E_{\rm b}$ is the binding energy of the species, $T_{\rm  dust}$ is the dust temperature, and $\nu_{\rm s}$ is the characteristic vibrational frequency of the species,
\begin{equation}
\nu_{\rm s} = \sqrt{\frac{ 2 N_{\rm s} k_{\rm B} E_{\rm b}}{\pi^2 \mu m_{\rm H}}} \quad \mathrm{s}^{-1} .
\end{equation}
Here, $N_{\rm s}$ is the number density of surface site (assumed to be 1.5$\times$10$^{15}$ site cm$^{-2}$).

Cosmic-rays and photons also desorb species from grains. The cosmic-ray desorption rate is calculated using the formalism of \citet{hasegawa1993}.
\begin{equation}
k_{\rm crd} = f(70 \mathrm{K})\, k_{\rm td}(70 \mathrm{K})\, \frac{\zeta_{\rm  CR}}{5.0 \times 10^{-17} \mathrm{s}^{-1}} \quad \mathrm{s}^{-1}
\end{equation}
where $k_{\rm td}$(70K) is the thermal desorption rate at 70~K and $f(70 \mathrm{K})$ is the ratio of the grain cooling timescale via desorption of species to the timescale of subsequent heating events. We adopt $f(70 \mathrm{K})$ as 3.16 $\times 10^{-19}$ for the grain size of  0.1~$\mu$m from  \citet{hasegawa1993}.  

 The photodesorption rate by UV photons is calculated following the method of \citet{Woit09}:
\begin{eqnarray}
\label{eq:PDrate}
k_{\rm phd}  &=&  \sigma_{\rm gr} \frac{ n_{\rm gr}}{N_{\rm p} \times n_{\rm act}} Y G_{\rm  dust} F_{\rm D} \quad \mathrm{s}^{-1}\quad  if\quad N_{\rm m} < N_{\rm p} \\ \nonumber
            &=&  \sigma_{\rm gr} \frac{ n_{\rm gr}}{n_{\rm ice}} Y G_{\rm  dust} F_{\rm D}      \quad \mathrm{s}^{-1} \quad\quad     if\quad N_{\rm m} \geq N_{\rm p}
\end{eqnarray}
where $n_{\rm act}$ (=$ 4\pi\, a_{\rm gr}^2\, N_s \,n_{\rm gr}$) is the number of active surface places in a monolayer of ice mantle per volume, $n_{\rm ice} $(=$\sum_j n_{\rm s}(j)$) is the total number of ice species, and $Y$ is the photodesorption yield (the number of ice species ejected per incident photon). $F_{\rm D}$ is the conversion factor of $G_{\rm  dust}$ to the photon number flux, which is 1.93~$\times$~10$^8$ cm$^{-2}$ s$^{-1}$ for the Draine field and  2.33~$\times$~10$^{8}$ cm$^{-2}$ s$^{-1}$ for BB1.0. $N_{\rm m}$ (= $n_{\rm ice} / n_{\rm act}$) is the number of monolayers. We assume $N_{\rm p}$~=~2 because the photodesorption by UV photons occurs in the upper $\sim$2 monolayers \citep{oberg2007}.
 
%
%
 We follow the model of H$_2$ \, formation on interstellar dust grains via physisorption and chemisorption from \citet{cazaux2002, cazaux2004, cazaux2010erratum}.
\begin{equation}
\label{eq:h2form}
k_{\mathrm{H}_2}=\frac{1}{2}n_{\rm  H} v_{\rm  H} n_{\rm gr} \sigma_{\rm gr} \epsilon_{{\rm H}_2}S_{{\rm H}}\, \quad \mathrm{cm}^{3}\, \mathrm{s}^{-1}, 
\end{equation}
where $n_{\rm  H}$ and $v_{\rm  H}\,\,( = 1.45\times 10^{4}\, \sqrt{T_{\rm  gas}}$ cm s$^{-1}$) are the number density and thermal velocity of H atoms in the gas phase, and $S_\mathrm{H}$ is the sticking coefficient of the H atoms \citep{HM79},
\begin{equation}\label{sticking}
S_\mathrm{H}=\left(1+0.04\left(\frac{T_{\rm  gas}+T_{\rm  dust}}{100}\right)^{1/2}+0.2\frac{T_{\rm  gas}}{100}+0.08\left(\frac{T_{\rm  gas}}{100}\right)^2\right)^{-1}   .
\end{equation}
The formation efficiency $\epsilon_{\mathrm{H}_2}$ is given by \citet{cazaux2002,cazaux2004,cazaux2010erratum}:
\begin{align}
\label{eq:h2form2}
\epsilon_{{\rm H}_2}& =\left({\rm A}+1+{\rm B}\right)^{-1}\xi \nonumber \\
\epsilon_{{\rm H}_2}& =\left(\frac{\mu F}{2 \beta_{{\rm H}_2}}+1+\frac{\beta_{{\rm H}_P}}{\alpha_{{\rm P}_C}}\right)^{-1}\xi\, . 
\end{align}
 We set $A$ to zero to make newly formed H$_2$ molecules  leave very cold dust surfaces, which is equivalent to the equation (13) in \citet{cazaux2002}.

   We include the electron attachment to grains and the recombination of cations with the negatively charged grains adopted from the Ohio State University Astrophysical Chemistry Group gas-phase database \citep{smith2004}.  The initial abundances in our model are listed in Table~\ref{initabun}, which represent the molecular cloud abundances approximated from \citet{aikawa99}.

%
%

\subsection{Gas energetics}  \label{sec:gasenergetics}
To obtain the gas temperature, the steady state thermal balance should be solved. We consider only important heating and cooling processes:

\noindent
 \textit{Photoelectric heating and recombination cooling by PAHs and grains :} FUV photons absorbed by PAHs and grains create energetic (several eV) electrons to heat the gas. For this heating rate, \citet{WD01PEH} provide an approximated formula for the recent grain size distribution models \citep{WD01},
 \begin{eqnarray}
 \label{eq:PEWD}
\Gamma_\mathrm{PE} = 10^{-26} G_{\rm  dust} n \frac{1.84 + 3.81 T_{\rm  gas}^{0.089}}{1+0.08348\psi^{0.328}[1+0.00391\psi^{0.778}]} \quad \mathrm{erg}\,\mathrm{s}^{-1}\,\mathrm{cm}^{-3},
\end{eqnarray}
with $\psi=(G_{\rm  dust}\sqrt{T_{\rm  gas}})/n_{\rm e}$. Where $n$ (= $n_{\rm H} + 2n_{\rm H2}$) is the total hydrogen number density, $T_{\rm  gas}$ is the gas temperature, $n_{\rm e}$ is the electron number density, and $G_{\rm  dust}$ is the dust attenuated FUV strength described in Sec. \ref{sec:fuvrt}. We use the 18th model ($R_{\rm V}$~=~5.5) in Table 2 in \citet{WD01PEH}. This approximation is valid in the range of 10~K $\leq$  $T_{\rm  gas} \leq$ 10$^4$~K and  10$^2$\,~K$^{1/2}$~cm$^3  \leq \psi \leq\, 10^5$~K$^{1/2}$~cm$^3$, and it can be extended to $\psi \leq 10^2$\,~K$^{1/2}$~cm$^3$. 

The recombination cooling is approximated by 
\begin{eqnarray}
\label{eq:recomb}
\Lambda_\mathrm{RC}&=&10^{-28}{\rm erg}\,{\rm s}^{-1} \, {\rm cm}^{-3}\times n_{\rm e} n\,\,T_{\rm gas}^{0.4440 + 2.067/x_{\psi}} \\
                   &\times& \exp\left({-7.806 +1.687x_{\psi}-0.06251 x_{\psi}^2}\right)\quad \mathrm{erg}\,\mathrm{s}^{-1}\,\mathrm{cm}^{-3},\nonumber
\end{eqnarray}
where $x_{\psi}=\ln{\psi}$. This equation works when the gas temperature is higher than 10$^3$~K, and it is fairly accurate when $10^2 \, \mathrm{K}^{1/2}\mathrm{cm}^3\le \psi\le 10^6\, \mathrm{K}^{1/2}\mathrm{cm}^3$. If $\psi$ is out of this range, we use the constant value of $\Lambda_\mathrm{RC}/n_{\rm e} n$ at $\psi = 10^2\,\mathrm{K}^{1/2}\mathrm{cm}^3$ and 10$^6\, \mathrm{K}^{1/2}\mathrm{cm}^3$ (see \citet{Rollig13}).
 
 \citet{Spaans94} calculated the photoelectric heating rate for the blackbody radiation field of a effective temperature ($T_{\rm  eff}$). 
 As photons below 6~eV also contribute the photoelectric heating, they used a modified unit, $G_{\rm dust}^{'}$ normalized to the Habing field to have the same integrated intensity from \textit{2~eV to 13.6~eV}. 
  The heating rate can be calculated by multiplying Eq.\ref{eq:PEWD} (with $G_{\rm dust}^{'}$) by a simple correction factor, $e(T_{\rm  eff})$,
 \begin{equation}\label{eq:eff}
e(T_{\rm  eff})=\left(\frac{T_{\rm  eff}}{30,000 \mathrm{K}} \right)\times \left[\mathrm{log} \left(1.4\times 10^{-4}\, \psi'\right)\right]^{s(T_{\rm  eff})}
\end{equation}
where $s(T_{\rm  eff})$ = -1 if $T_{\rm  eff} <$~20,000~K and $\psi^{'} = (G_{\rm  dust}^{'}\sqrt{T_{\rm  gas}})/n_{\rm e} > 2 \times 10^4 \mathrm{K}^{1/2} \mathrm{cm}^3$, and 0 otherwise. 
 In order to describe the photoelectric heating rate for a given $G_{\rm dust}$ normalized to the Habing field to have the same integrated intensity from \textit{6~eV to 13.6~eV}, $G_{\rm dust}^{'}$ is  corrected as,
\begin{equation}
G_{\rm  dust}^{''} = G_{\rm  dust}^{'}\frac{\int_{912}^{6196} BB(T_{\rm  eff}) d\lambda }{\int_{912}^{2050} BB(T_{\rm  eff}) d\lambda}
				/ \frac{\int_{912}^{6196} BB(30,000 \mathrm{K}) d\lambda }{\int_{912}^{2050} BB(30,000 \mathrm{K}) d\lambda}
\end{equation} 
where $BB(T_{\rm  eff})$ is the Planck function with $T_{\rm  eff}$. 
The corrected unit, $G_{\rm  dust}^{''} = 6.67 G_{\rm  dust}^{'}$ for the model of BB1.0 is used in Eqs.~\ref{eq:PEWD} --\ref{eq:eff}. 

\begin{figure*}
    \includegraphics[width=0.8 \textwidth]{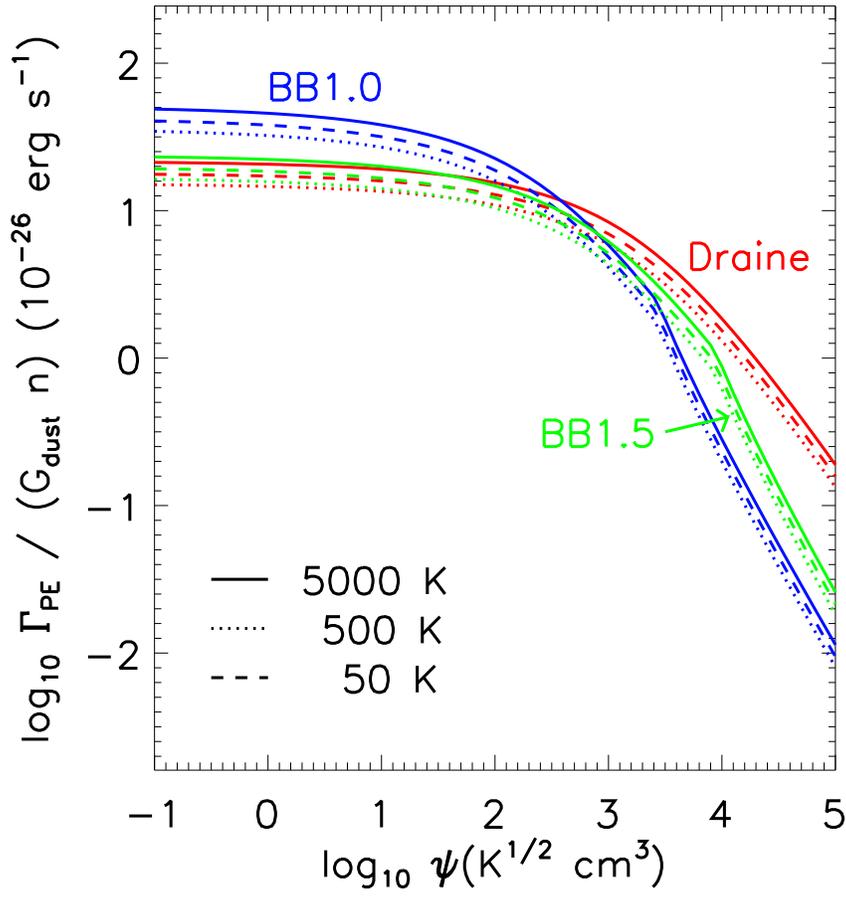}
    \caption{ Photoelectric heating rates of the Draine field, BB1.0, and BB1.5 (see text).}\label{fig:pebb} 
\end{figure*}

Fig.~\ref{fig:pebb} shows photoelectric heating efficiencies as functions of $\psi$ (in the unit of $G_{\rm dust}$) for given spectral types. Higher energy (shorter wavelength) photons photoelectrically heat the gas more efficiently. For $\psi > 10^3 \, \mathrm{K}^{1/2} \mathrm{cm}^3$, most grains are neutral and positively ionized, and only high energy photons ($>$ 6~eV) can remove electrons from the grains. Therefore, the efficiency of the Draine field is higher than that of BB1.0. 

However, for $\psi < 10^3 \, \mathrm{K}^{1/2} \mathrm{cm}^3$, a large portion of a grain is negatively charged (see Fig. 9 in \citealt{WD01PEH}).
 As a result, photons with energies lower than 6~eV can remove electrons from the negatively charged grains, which have the first electron affinity lower than 6~eV \citep{Bakes94,WD01PEH}.
The integrated intensity from 2~eV to 6~eV is larger  than that from 6~eV to 13.6~eV by a factor of $\sim$6 for BB1.0, while most of the intensity is deposited above 6~eV for the Draine field. This results in a higher efficiency for BB1.0 than for the Draine field for a given $G_{\rm dust}$.
\noindent


 \textit{$H_2$ vibrational heating :} The gas is heated if a hydrogen molecule excited by FUV radiation is collisionally de-excited. \citet{Rollig06} provides an approximated formula for the Draine field. To apply the other UV radiation field, we assume that the pumping and dissociation rates are proportional to the local H$_2$ photodissociation rate ($k_{\rm ph}^{\rm H_2}$), which  accounts for H$_2$ self-shielding (Eq. \ref{eq:H2-shield}) as well as the attenuation of the FUV radiation field  as described in Sec. \ref{sec:chem}. Then, the modified equations is,
 \begin{eqnarray}\label{eq:H2heat}
  \Gamma_{\mathrm H_2^\star}&=&n_{\mathrm H_2}\,\frac{1.8\times 10^{-11}\,k_{\rm ph}^{\rm H_2}}{1+\left(\frac{1.9\times 10^{-6}+ 9.1\,k_{\rm ph}^{\rm H_2}}{\gamma_c \, n}\right)}\quad \mathrm{erg}\,\mathrm{s}^{-1}\,\mathrm{cm}^{-3},
 \end{eqnarray}
 with the collision rate, $\gamma_c = 5.4\times 10^{-13}\,\sqrt{T_{\rm  gas}}$~cm$^{-3}$~s$^{-1}$.

\noindent

 \textit{$H_2$ formation heating :} If we assume that each H$_2$ formation process releases 1/3 of its binding energy to heat the gas, the corresponding heating rate (\citealt{Rollig06})  is
\begin{equation}\label{rate_form}
\Gamma_{\rm  form}=2.4\times10^{-12}\,k_{\rm  H_2}\,n_{\rm H} \quad \mathrm{erg}\,\mathrm{s}^{-1}\,\mathrm{cm}^{-3},
\end{equation}
where the H$_2$ formation rate $k_{\rm  H_2}$ is  described in Eq. \ref{eq:h2form}.

\noindent

 \textit{H$_2$ dissociation heating :} About 10\% of the radiative decays in the H$_2$ dissociation deliver about 0.25 eV to the gas. This heating rate is taken from \citet{Meijerink05},
\begin{eqnarray}
\Gamma_{\rm  H_2}& =& 2.63 \times 10^{-13}\ n_{\rm H_2}\ k_{\rm ph}^{\rm H_2}\quad\mathrm{erg}\,\mathrm{s}^{-1}\,\mathrm{cm}^{-3}.
\end{eqnarray}
 We also slightly modify the equation with the local H$_2$ photodissociation rate ($k_{\rm ph}^{\rm H_2}$).
 
\noindent
\textit{ C ionization heating :} When a neutral carbon is ionized, a photo-electron released with the energy around 1~eV heats the gas \citep{Woit09} with the rate of
\begin{equation}
\Gamma_{\rm C} = 1.602 \times 10^{-12}\, n_{\rm C}\, k_{\rm ph}^{\rm C}\quad\mathrm{erg}\,\mathrm{s}^{-1}\,\mathrm{cm}^{-3},
\end{equation}
where $n_{\rm C}$ is the neutral carbon number density and $k_{\rm ph}^{\rm C}$ is the local carbon photoionization rate corrected by the shielding factor in Eq.~\ref{eq:beta_c}.

\noindent
 \textit{Cosmic ray heating :} For the low degree of ionization, $< 10^{-4}$, the primary ionization by a cosmic ray particle releases the energy of about 9 eV to heat the gas. The heating rate is $\Gamma_{\rm  CR}= 1.5 \times 10^{-11} \, \zeta_{\rm CR} \, n$~ erg cm$^{-3}$ s$^{-1}$.

%
%

\noindent

\textit{Fine structure line cooling :} The most prominent forbidden fine structure lines at the surface of outflow cavity walls are [OI] 63 $\mu$m, [OI] 146 $\mu$m, [CI] 369 $\mu$m, [CI] 609 $\mu$m, [SiII]~34.8~$\mu$m, and [CII] 158 $\mu$m. We calculate the cooling rate using the escape probability method (e.g. \citealt{Tielens05}) and use the atomic and cationic data taken from the Leiden Atomic and Molecular Database \citep[LAMBDA]{Scho05} except for Si$^+$ \citep{HM79}. The column densities of these species are assumed to be the products of the distance to the nearest boundary from the current grid and the local number densities of those.
\noindent

\textit{H$_2$ vibrational cooling :} Vibrational lines of H$_2$ can contribute to the cooling of the gas. Due to the large energy gap (6000 K) between the ground state and the first excited state, we use the two level approximation given in \citet{Rollig06},
\begin{eqnarray}
 \label{h2cool}
 \Lambda_{\mathrm H_2}&=&n\,n_{\mathrm H_2}\,9.1\times 10^{-13}
\,\gamma_c\,\exp(-6592\,{\rm K}/T_{\rm  gas}) \nonumber
 \\
 & &\times\frac{8.6\times 10^{-7}+0.48\, k_{\rm ph}^{\rm H_2} }{\gamma_c\,n+8.6\times 10^{-7}+0.48\,k_{\rm ph}^{\rm H_2}}\quad \mathrm{erg}\,\mathrm{s}^{-1}\,\mathrm{cm}^{-3}
 \end{eqnarray}
  where $k_{\rm ph}^{\rm H_2}$ and $\gamma_c$ are described in Eq. \ref{eq:H2heat}.
 
\noindent

\textit{Gas-grain cooling/heating :} The temperature difference between gas and dust leads to the transfer of heat. This can be an important coolant near the surface of the dense PDR where $T_{\rm  dust} < T_{\rm  gas}$. The rates are proportional to $T_{\rm  dust} - T_{\rm  gas}$. We adopt the results of \citet{Burke1983} with the dust cross section per H nucleus of $\sigma_{\rm d} = 2.98\times 10^{-21} \mathrm{cm}^{-2}$ \citep{Rollig13},
\begin{eqnarray}
\label{eq:ggcool}
\Gamma_{\rm  coll.} &=& 4.4\times 10^{-33} n^2 \sqrt{T_{\rm  gas}}\left(\frac{\sigma_{\rm d}}{2.98\times 10^{-21} \mathrm{cm}^{-2}}\right)\\ \nonumber
                   & &\times[1-0.8\exp(-75/T_{\rm  gas})](T_{\rm  dust} - T_{\rm  gas})\quad \mathrm{erg}\,\mathrm{s}^{-1}\,\mathrm{cm}^{-3}.
\end{eqnarray} 
%
%
\noindent
\textit{Molecular cooling by CO and H$_2$O :} If CO and H$_2$O molecules exist, their lines can provide more efficient cooling than [OI] and [CII] lines.  We calculate the molecular line cooling rate following the method of \citet{Meijerink05} and \citet{Yan97}, which used the fitted cooling rate coefficients of \citet{Neufeld93} and \citet{Neufeld95}. Isotope ratios are assumed to be ${}^{12}$C~/~${}^{13}$C~=~69 and  ${}^{16}$O~/~${}^{18}$O~=~557 (Wilson 1999). The column densities of CO and H$_2$O are calculated by the same methods as used for the column densities of atoms in the fine structure line cooling.

\noindent
\textit{Ly~$\alpha$ and OI-6300~\AA \,cooling :} At a high gas temperature, Lyman $\alpha$ and OI-6300~\AA\, line cooling are important cooling processes. We adopt simple approximated formulae from \citet{Sternberg89}:
\begin{equation}
\Lambda_{\rm Ly \alpha} = 7.3 \times 10^{-19}\, n_{\rm H}\, n_{\rm e} \,\exp\left(-118\,400/T_{\rm gas}\right)\quad \mathrm{erg}\,\mathrm{s}^{-1}\,\mathrm{cm}^{-3}
\end{equation}
and 
\begin{equation}
\Lambda_{\rm OI-6300} = 1.8 \times 10^{-24}\, n_{\rm O} \,n_{\rm e}\, \exp\left(-22\,800/T_{\rm gas}\right) \quad \mathrm{erg}\,\mathrm{s}^{-1}\,\mathrm{cm}^{-3}
\end{equation}
with the atomic oxygen number density, $n_{\rm O}$.

\subsection{Line radiative transfer}
 We have developed a new solver of a non-LTE line Radiative transfer In general Grid (RIG). RIG has been upgraded from RATRAN \citep{Hogherheijde00} and use the same ray tracing method described in Sec. \ref{sec:raytrace}.
 
This code solves the equation of radiative transfer and the equation of statistical equilibrium iteratively.  When a photon propagates with a distance ($ds$), the intensity ($I_{\rm \nu}$) at a frequency of $\nu$ varies  as 
 \begin{equation}
 \frac{d I_{\rm \nu}}{ds} = j_{\rm \nu} - \alpha_{\rm \nu} I_{\rm \nu},
 \end{equation}
 where $j_{\rm \nu}$ and $\alpha_{\rm \nu}$ are the local emission and absorption coefficients, respectively. These coefficients are related with the properties of molecules and dust particles.

For molecular radiation, each transition has the two coefficients as
 \begin{eqnarray}
 j_{\rm \nu}^i (gas) & = & \frac{h\nu_i}{4\pi}n_u A_{\rm ul}\phi_i(\nu) \\
 \alpha_{\rm \nu}^i (gas) & = &  \frac{h\nu_i}{4\pi}\left(n_l B_{\rm lu} - n_u B_{\rm ul}\right)\phi_i(\nu),
 \end{eqnarray}
 where $A_{\rm ul}$, $B_{\rm ul}$, and $B_{\rm lu}$  are the Einstein coefficients. $n_{\rm l}$ and $n_{\rm u}$ are lower and upper level populations, respectively. h$\nu_i$ is the energy difference between the lower and upper levels. The line profile is assumed to be a Doppler profile:
 \begin{equation}
 \phi_i(\nu) =\frac{1}{\sigma \sqrt{\pi}}exp\left[-\left(\nu - \nu_i - \vec{v}\cdot\vec{n}\frac{\nu_i}{c}\right)^2 / \sigma^2\right], 
 \end{equation}
 where $\sigma$ is the Doppler width and $\nu_i$ is the center frequency of the transition, $\vec{v}$  is the local velocity field, and $\vec{n}$ is the direction vector of the photon-propagation. Our code considers line overlap in complex molecules. The two coefficients for the molecules are
 \begin{eqnarray}
  j_{\rm \nu} (gas) & = & \sum j_{\rm \nu}^i (gas) \\
 \alpha_{\rm \nu} (gas) & = & \sum \alpha_{\rm \nu}^i (gas).
 \end{eqnarray}

For dust continuum radiation, the two coefficients are 
\begin{eqnarray}
j_{\rm \nu} (dust) & = & \alpha_{\rm \nu} (dust) B_{\rm \nu} (T_{\rm  dust}) \\
\alpha_{\rm \nu} (dust) & = & k_{\rm \nu} \rho_{\rm  dust},
\end{eqnarray}
where B$_{\rm \nu}$ is the Planck function for a given dust temperature. k$_{\rm \nu}$ and $\rho_{\rm \rm dust}$ are the dust opacity and density, respectively.

 When we calculate the local radiation field, we determine the level populations through the equation of statistical equilibrium:
\begin{equation}
n_l \left[\sum_{k < l} A_{\rm lk} + \sum_{k \neq l} \left(B_{\rm lk}\bar{J_{\rm lk}} + C_{\rm lk}\right)\right] = 
\sum_{k > l}  n_k A_{\rm kl} + \sum_{k \neq l } n_k \left(B_{\rm kl}\bar{J_{\rm lk}} + C_{\rm kl}\right),
\end{equation}
where $C_{\rm kl}$ ($C_{\rm lk}$) is the collision rate from level $k~(l)$ to $l~(k)$ and $\bar{J_{\rm lk}}$ is
\begin{equation}
\bar{J_{\rm lk}} \equiv \int d\Omega \int d\nu\, I_{\rm \nu}\, \phi_{\rm lk}(\nu).
\end{equation}
 
 RATRAN solves the line radiative transfer using an accelerated Monte-Carlo method, which splits $\bar{J_{\rm lk}}$ into a local contribution and  an external field,
\begin{equation}
\bar{J_{\rm lk}} = \bar{J_{\rm lk}}^{external}  + \bar{J_{\rm lk}}^{local} 
\end{equation}
or, in a view of model photons,
\begin{eqnarray}
\bar{J_{\rm lk}} & = & \left[\sum_{i} I_{i}^{ext} e^{-\tau_i}\phi_{\rm lk}(\nu_i) + \sum_{i} S_{\rm \nu_i} 
\left( 1 -   e^{-\tau_i}\right) \phi_{\rm lk}(\nu_i)\right] / \sum_{i} \phi_{\rm lk}(\nu_i).
\end{eqnarray}
$I_{i}^{ext}$  is the intensity entering into the local cell, $\tau_i$ (the local optical depth) and $S_{\rm \nu_i}$ (the local source function)  are given as,
\begin{eqnarray}
\tau_i & = & \left(\alpha_{\rm \nu_i} (gas) + \alpha_{\rm \nu_i} (dust)\right) \cdot ds \\
S_{\rm \nu_i} & = & \frac {j_{\rm \nu_i} (gas) + j_{\rm \nu_i} (dust)}{\alpha_{\rm \nu_i} (gas) + \alpha_{\rm \nu_i} (dust)}
\end{eqnarray}

RATRAN finds a local solution for a grid by solving the equation of the statistical equilibrium and the local radiation field for the given external radiation field, then finds a global solution for all grids. We have upgraded the local solution finding method with ``newt" subroutine \citep{press92} in RIG, which can cope with line overlaps among multiple molecular and atomic species. 
We also update ``SKY" in RATRAN to make a spectral image in the general coordinate for  any given inclination.

%
%
\section{ Benchmarking }\label{sec:benchmark}

\subsection{Non-LTE line radiative transfer in the ($r$, $\delta$) coordinates}\label{sec:bench_rig}
\begin{figure*}
\includegraphics[width=0.9 \textwidth]{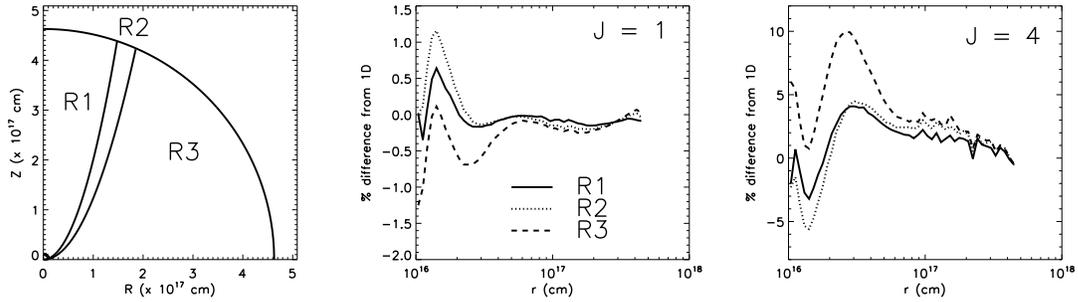}
\caption {The comparisons between 1D RATRAN and RIG. The spherical model adopted for comparisons is an analytic inside-out model \citep{shu77} of B335, and R1, R2, and R3 in the left panel represent three different $\delta$ layers. 
(see Sec.~\ref{sec:bench_rig}). } \label{fig:rig_bench}
\end{figure*}

 In order to confirm that the ray-tracing scheme in the ($r$, $\delta$) coordinates is reliable, we compare RIG with the ($r$, $\delta$) coordinate system with 1D RATRAN. 
For this test, we run  the bench mark test of model 2b in \citet{van Zadelhoff2002}\footnote{http://www.strw.leidenuniv.nl/astrochem/radtrans/}. This is an analytical inside-out collapse model \citep{shu77} of B335 for an optically thick case with a constant abundance of HCO$^+$ of $1\times 10^{-8}$. The benchmark test shows that the differences among participating models are 2\% and 20\% in $J$=1 and $J$=4, respectively \citep{van Zadelhoff2002}.
 For this test, we divide the envelope with the spherical symmetric density structure into three $\delta$ regions: R1, R2, and R3 as shown in the left panel of  Fig.~\ref{fig:rig_bench}. The three regions should have the same level population at the same radius  because the envelope has a spherical symmetric density structure. The differences of the level populations are smaller than those in the benchmark test by a factor of two. Therefore, the ray-tracing in the ($r$, $\delta$) coordinates is trustworthy, and these coordinates  can be used in non-LTE line radiative transfer and FUV radiative transfer.
 
\subsection{FUV radiative transfer in PDR}\label{sec:fuv_benchmark}
 The FUV radiative transfer through the dusty material could be adopted directly from the result of RADMC-3D. However, in order to have a high spatial resolution in the outflow cavity walls, a large number of grids is required in RADMC-3D, resulting in a great increase of computational time. Therefore, we have developed our own less time-intensive code for the FUV radiative transfer in the ($r$, $\delta$) coordinates as described in the Sec.~\ref{sec:fuvrt}.

\begin{figure*}
\includegraphics[width=0.9 \textwidth]{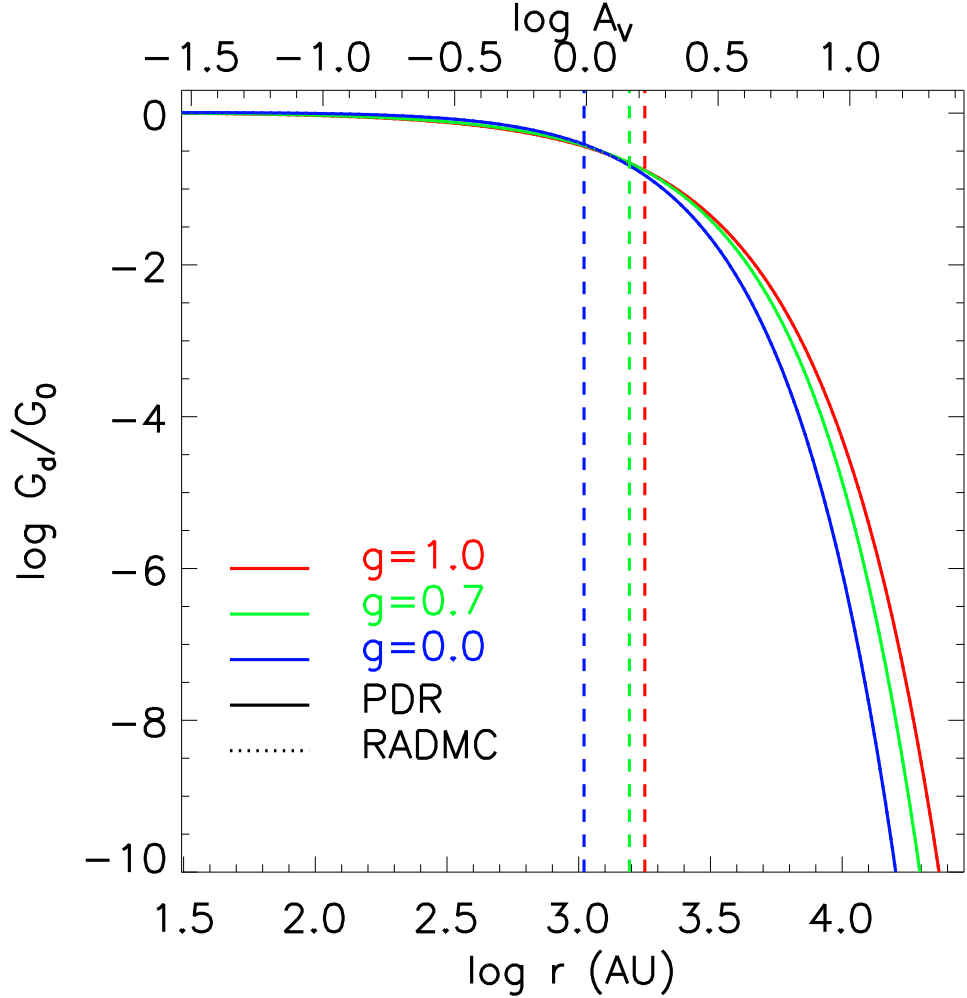}
\caption {The dust attenuated FUV strength normalized to the incident FUV strength for the benchmark test of FUV radiative transfer (see Sec.~\ref{sec:fuv_benchmark}). A central protostar is only a UV source, and an envelope has a constant density of 10$^5$ cm$^{-3}$. Solid and dotted  lines indicate the result of our model and RADMC, respectively. Red, green, and blue lines represent results with the mean scattering angle, g=1.0, g=0.7, and g=0.0, respectively. The average visual extinction of unity ($\left<A_{\rm V}\right>$ =1) for the three models are plotted as vertical dashed lines.  The horizontal axis on the top represents the 1D visual extinction.} \label{fig:fuv_benchmark}
\end{figure*}

 In order to confirm that our calculation of the FUV radiative transfer is reasonable, a simple spherical model with a constant density of 10$^5$ cm$^{-3}$ is tested using our code and RADMC-3D.
In this comparison, the only source of FUV radiation is the central protostar. In the FUV radiative transfer, scattering (as well as absorption) by dust grains must be considered; three types of scattering with g=1.0 (pure forward scattering), g=0.0 (isotropic scattering), and g=0.7 (mean scattering angle in the UV range) are tested in the comparison as seen in Fig.~\ref{fig:fuv_benchmark}.
The two codes used the same 100 grid cells from 30 AU to 30,000 AU.
Fig.~\ref{fig:fuv_benchmark} shows that the results by two codes are very consistent. 
$A_{\rm V}$ on top axis is derived from the relation between $A_{\rm V}$ and the column density of hydrogen. The average visual extinction $\left<A_{\rm V}\right>$, which is derived by Eq. \ref{eq:Av}, is higher than $A_{\rm V}$ as the scattering is more forward directed.
However, for the pure forward scattering (g=1.0), which has an analytic solution using the relation between $A_{\rm V}$ and the column density of hydrogen, $\left<A_{\rm V}\right>$ is only 1\% different from $A_{\rm V}$ when the albedo is considered. Therefore, our FUV radiative transfer code is reliable.

\subsection{Thermo-chemical part of PDR}
In order to test the reliability of our PDR code, we have run the four benchmark tests described in the PDR comparison study by R07: V1 ($n$~=~10$^3$~cm$^{-3}$ and $\chi$~=~10), V2 ($n$~=~10$^3$ cm$^{-3}$ and $\chi$ =~10$^{5}$), V3 ($n$~=~10$^{5.5}$~cm$^{-3}$ and $\chi$~=~ 10), and V4 ($n$~=~10$^{5.5}$~cm$^{-3}$ and $\chi$~=~10$^{5}$). These tests calculate the gas temperature and the chemistry self-consistently. A cloud with one dimensional slab geometry is assumed to be illuminated by an UV field in only one side. The same model parameters (Table~5 of R07), chemical species, and chemical reactions as those for the benchmark tests are used. As a result, we use the simple H$_2$ formation rate of $R_{\rm  H_2} = 3 \times 10 ^{-18} \sqrt{T_{\rm  gas}}\,n\, n_{\rm  H}$ instead of Eq. \ref{eq:h2form} and the formula of \citet{Bakes94} instead of Eq. \ref{eq:PEWD} and \ref{eq:recomb} for the photoelectric heating and the recombination cooling in the benchmark test. 
 The number of total grid cells is 300, and they are equidistant on the logarithmic scale of $A_{\rm V}$ between 10$^{-6}$ and 10.
For this test, we use the simple exponential form, exp($-$3.12~$A_{\rm V}$), for the dust attenuated FUV strength and the dust temperature  obtained from the analytical formula by \citet{Hollenbach1991}.  The chemistry is calculated until 10$^8$ yr to reach the steady state.

\begin{figure*}
    \includegraphics[width=0.495 \textwidth]{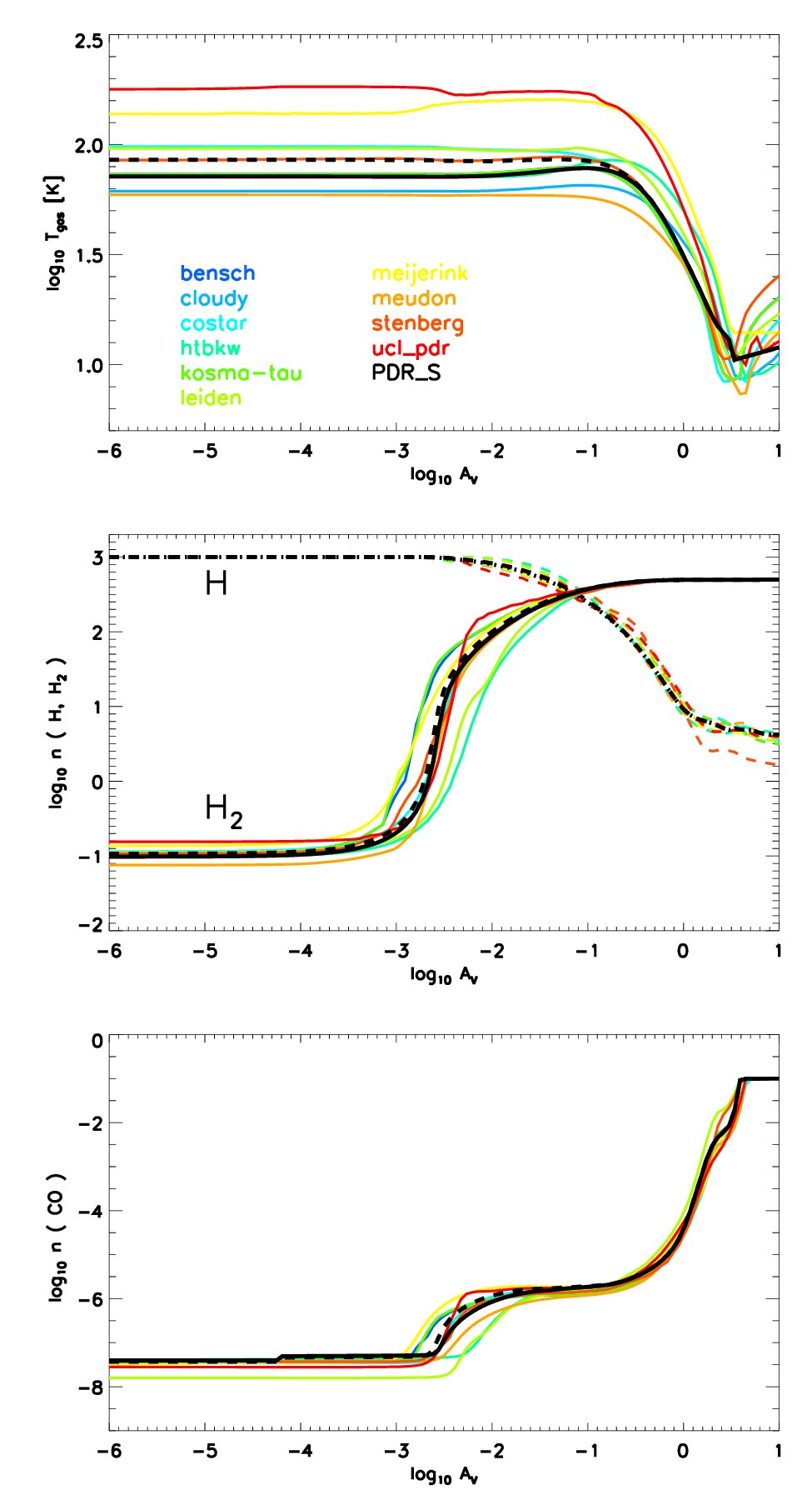}
    \includegraphics[width=0.495 \textwidth]{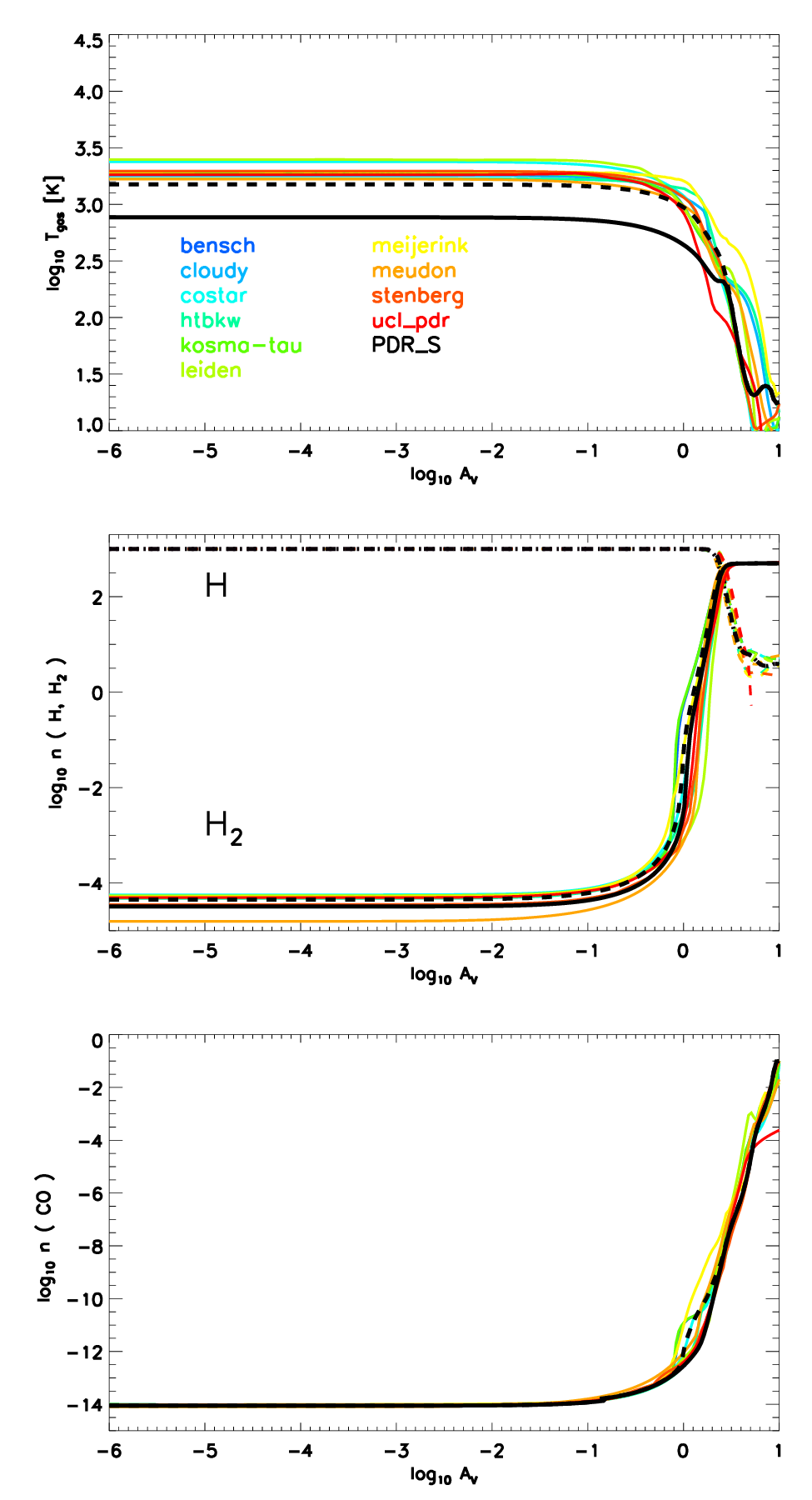}
        \caption{Benchmarking results of V1 (left, log~$n$~=~3 and log~$\chi$~=~1) and V2 (right, log~$n$~=~3 and log~$\chi$~=~5). Top, middle, and bottom rows show the gas temperature, the number densities of H and H$_2$, and the number density of CO, respectively. Color lines indicate different PDR model participating in the benchmark test (see text). PDR\_S represents our model. In V2, the model with the updated collision rate coefficients of O atom (solid black line) has lower gas temperature than the model with the collision rate coefficients of O atom used in the other models (dashed black line).}\label{fig:v1}
\end{figure*}

\begin{figure*}
    \includegraphics[width=0.495 \textwidth]{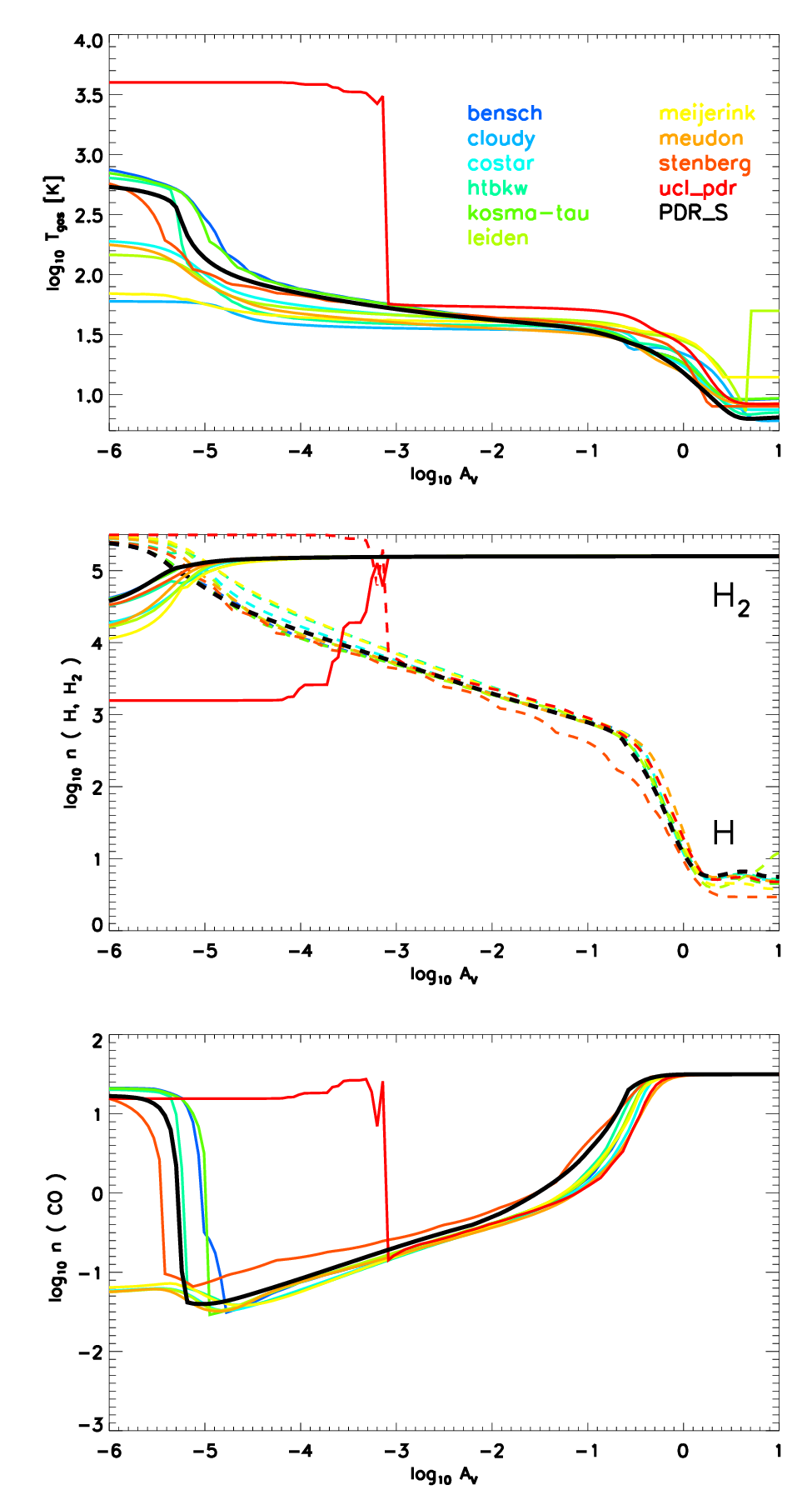}
    \includegraphics[width=0.495 \textwidth]{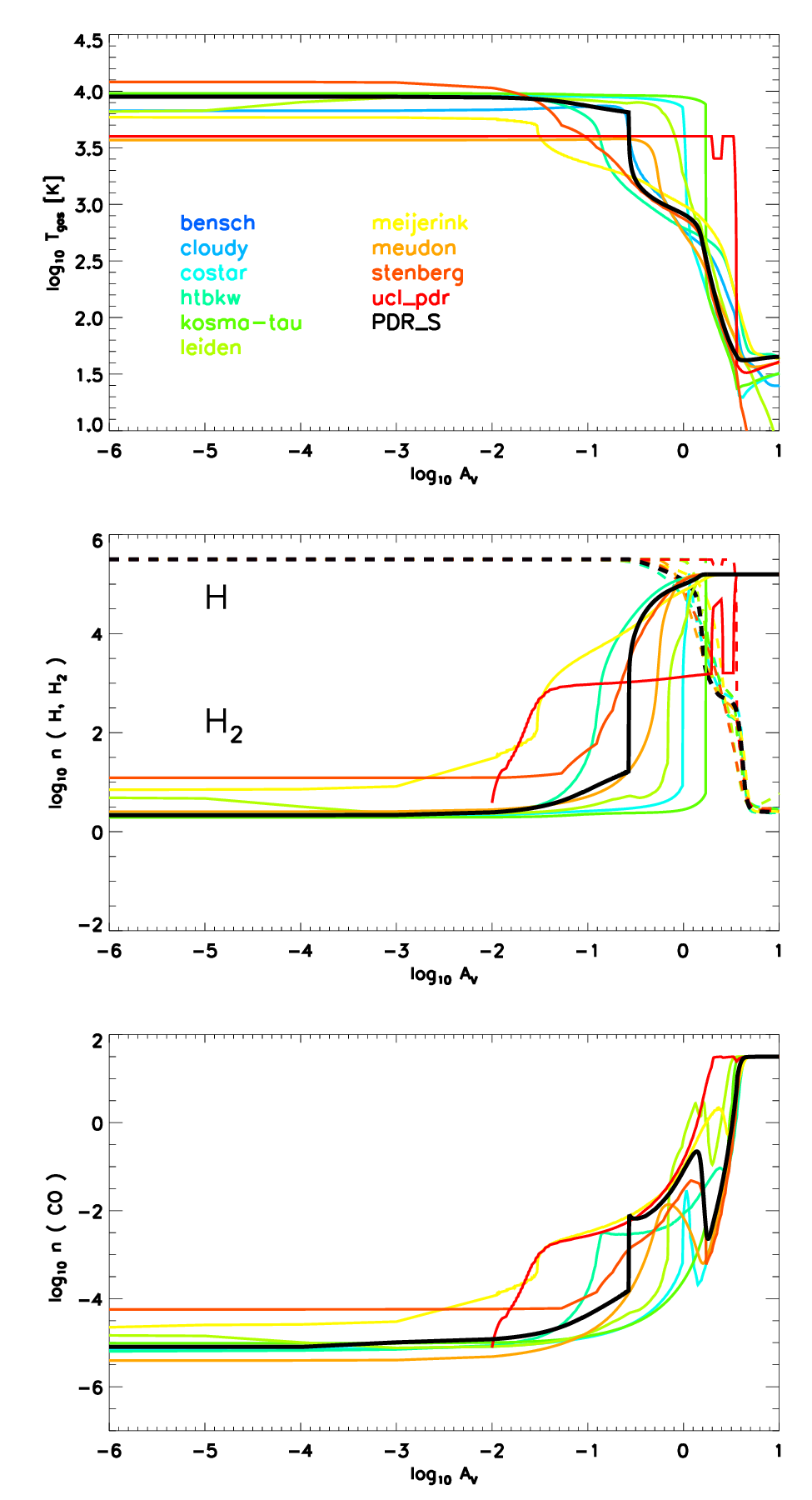}
        \caption{Benchmarking results of V3 (left, log~$n$~=~5.5 and log~$\chi$~=~1) and V4 (right, log~$n$~=~5.5 and log~$\chi$~=~5). }\label{fig:v3}
\end{figure*}

Figs. \ref{fig:v1} and \ref{fig:v3} show the results of our PDR model (PDR\_S) compared to those of other codes in R07\footnote{http://www.astro.uni-koeln.de/site/pdr-comparison/}: Cloudy \citep[e.g.][]{Abel05}, Costar \citep{Kamp01}, htbkw \citep[e.g.][]{Tiel85}, Kosma-tau \citep[e.g.][]{Rollig06}, leiden \citep[e.g.][]{Jans95}, Meijerink \citep{Meijerink05}, meudon \citep[e.g.][]{LePe04}, stenberg~\citep[e.g.][]{Sternberg89}, and ucl-pdr \citep[e.g.][]{Bell06}.  The overall agreement is very good, and the results of our PDR model fall within the scatter of the results produced by other codes. Therefore, our PDR model is reliable enough to be applied to more complicated models.

 The only notable difference between our model and others in R07 is the gas temperature of V2 model (see the right column in Fig. 1).  We use the updated collision rate coefficients of O atom with atomic hydrogen \citep{Abrahamsson2007}, which are larger than previous calculations by \citet{Launay1977} (used in other models) by a factor of 2-3 at temperature near 1000~K. Therefore, our V2 model has higher [OI] cooling rates resulting in lower gas temperatures in the lower $A_{\rm V}$.

\section{ 1D PDR model for warm CO  }\label{sec:co1d}
Before running a 2D model, we have made simple tests to check the PDR contribution to the FIR mid-\J\,($14\leq J\leq 24$) CO transitions with the 1D model. We have run the plane-parallel 1D model, similar to the benchmark tests, with our full chemistry and gas energetics. Though an approximated formula for the dust temperature in BB1.5 and BB1.0 is different from that in the Draine field \citep{Spaans94}, we use the same equation in the Section. \ref{sec:benchmark}. The explored parameter space is 2.0~$\leq$~log~$n \leq$~9.0 and 0.0 $\leq$~log~$G_0 \leq$~6.0 with a step of 0.5. 

\begin{figure*}
	\includegraphics[width=1.0 \textwidth]{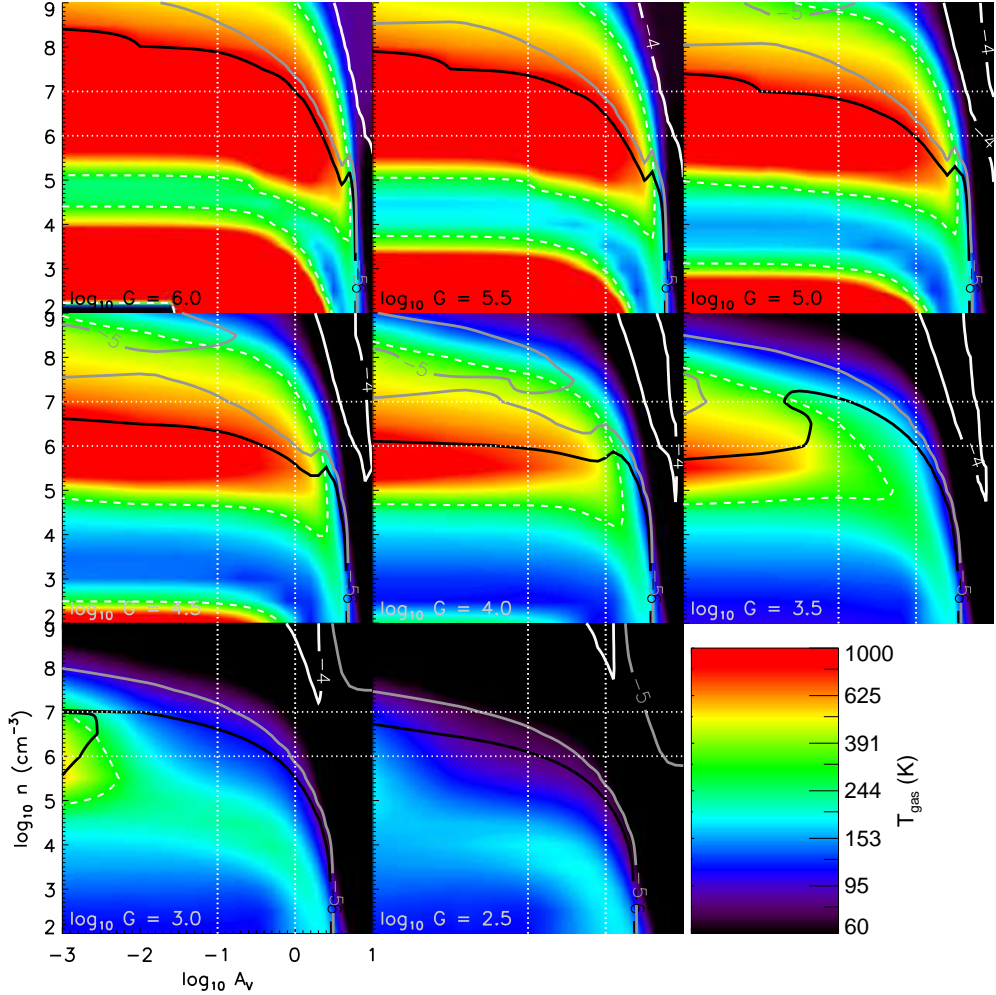}
	\caption{Gas temperature (image) and CO abundance (contour), as a function of visual extinction ($A_{\rm V}$) and the total hydrogen density ($n$), for a given FUV strength (presented inside boxes)  with the Draine field.
Black, grey, and white solid contours indicate the CO abundances of 10$^{-6}$, 10$^{-5}$, and 10$^{-4}$, respectively. White dashed contour represents the gas temperature of 300 K. Two vertical and horizontal dotted lines are the lines for  $A_{\rm V} = 0.1$ and $1$, and  ${\rm log}\, n\, = 6$ and $7$, respectively.
}\label{fig:draine_tk_nco}
\end{figure*}

\begin{figure*}
	\includegraphics[width=1.0 \textwidth]{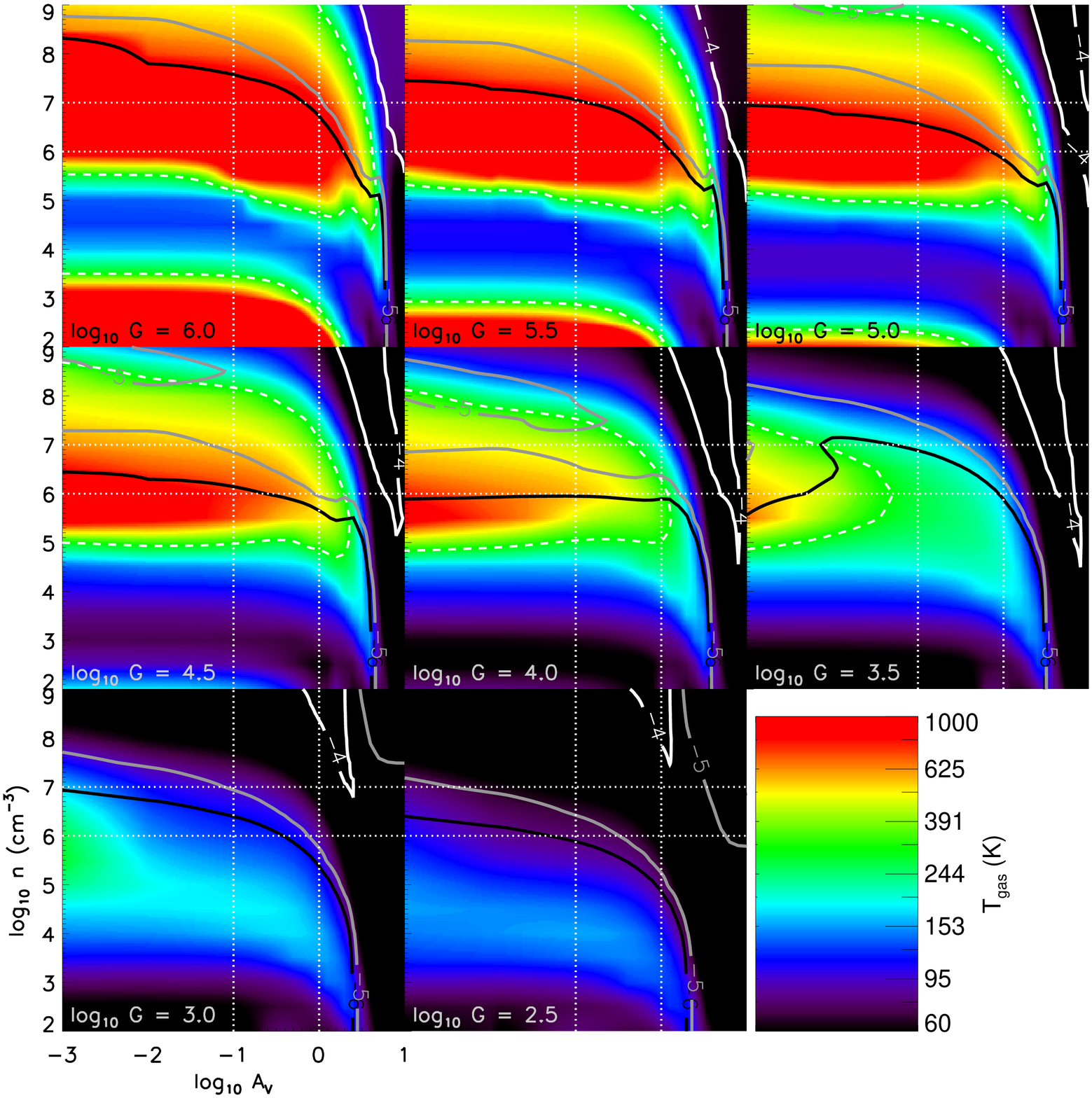}
	\caption{ The same as Fig. \ref{fig:draine_tk_nco} except for BB1.5.}\label{fig:bb1.5_tk_nco}
\end{figure*}

\begin{figure*}
	\includegraphics[width=1.0 \textwidth]{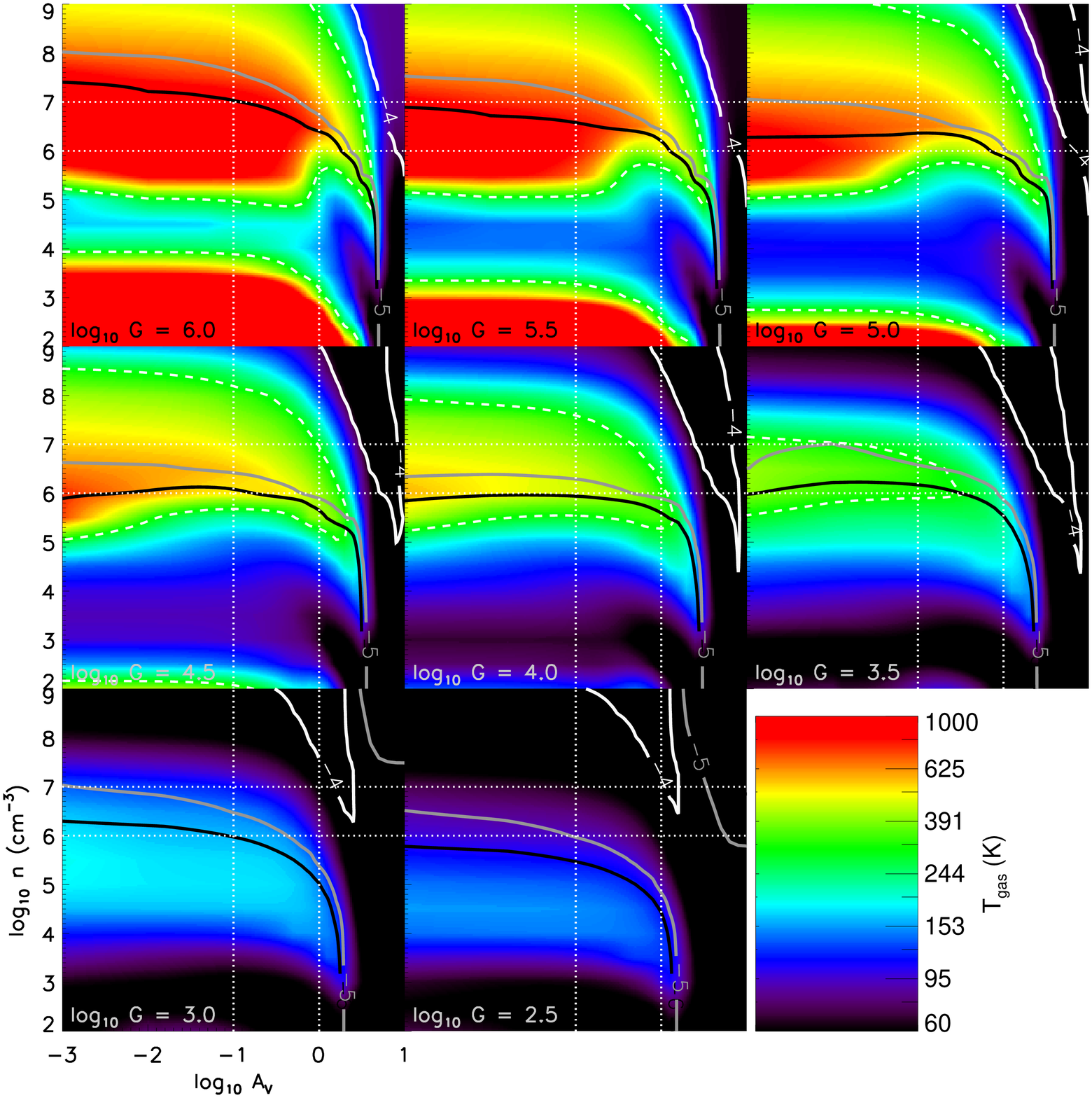}
	\caption{ The same as Fig. \ref{fig:draine_tk_nco} except for BB1.0.}\label{fig:bb1.0_tk_nco}
\end{figure*}

Figs.~\ref{fig:draine_tk_nco}-\ref{fig:bb1.0_tk_nco} show the gas temperature and CO abundance X(CO) in each physical point for the models with the Draine field, BB1.5, and BB1.0, respectively. The gas temperature is determined by the thermal balance between heating and cooling described in Sec.~\ref{sec:gasenergetics}.
As seen in Figs.~\ref{fig:draine_tk_nco}-\ref{fig:bb1.0_tk_nco}, the gas temperature is not a simple function of density. For log $G_0 >$4.0, the gas temperature has a dip at a range of density; for example, for log $G_0$ = 6, the temperature dip appears around $n=10^5$ cm$^{-3}$. This nonlinearity of temperature occurs because of the different dependence of density in heating and cooling.
In the physical conditions of test models, the dominant heating and cooling mechanisms are photoelectric heating by PAHs and small grains and the [OI] 63 $\mu$m emission line, respectively. 
The photoelectric heating rate is proportional to $n^p$ and $1<p<2$. However, the cooling rate by the [OI] line is proportional to $n^2$ and $n$ if the density is smaller and greater than the critical density ($\sim10^5 \mathrm{cm}^{-3}$) of the [OI] line, respectively (see Sec. 3.1 of \citet{kaufman99} for detailed explanations). As a result, these two combinations of different power indexes with density make the temperature dip, which appears at different densities depending on the UV spectral type and the UV strength 

As the FUV strength increases in the dense region (log $n \geq$ 6), the gas temperature also grows and more CO molecules are photodissociated near the surface. Interestingly, when log $G_0 \geq$~4, CO molecules survive even in the warm region with log X(CO) $\geq$~-5, which could emit the FIR mid-$J$ CO lines observed by \Herschel/PACS.  A high gas temperature enhances the CO formation rate to survive in this condition (see below). The models with BB1.5 and BB1.0 have slightly lower gas temperatures and higher CO abundances near the surface than the model with the Draine field.

 Because it is a simple 1D plane parallel model, we calculate the number of emitting CO molecules at the FIR mid-$J$ transitions with large velocity gradient code RADEX \citep{van der Tak2007}. We assume that the total hydrogen column density $N$(H) per visual extinction $A_{\rm V}$ is 1.87~$\times 10^{21}$~cm$^{-2}$, the column density of CO $N$(CO) at each $A_V$ position is the product of $N$(H) and the local CO abundance, and the line width is 1.0 km s$^{-1}$. Then the normalized level population in $J$ ($n(J)$; $\sum{n(J)} =1$) is calculated with RADEX. 
 
 The number of CO emitting in the $J$ level, $N(J)$, is defined as \begin{equation}\label{eq:nco}
 N(J) \simeq n(J) \times N(CO)\times \left[\frac{10^4}{G_0}\times (100 \mathrm{AU})^2 \right].
\end{equation}
The bracket is the area correction factor if the UV luminosity of the central source ($L_{\rm UV}$) is 0.1 $L_{\odot}$. At given $L_{\rm UV}$, the unattenuated FUV strength $G_0$ is approximated as: 
\begin{equation}
G_0 \simeq 10^4\frac{L_{\rm UV}}{0.1 \mathrm{L}_{\odot}}\left(\frac{r}{100\rm{AU}}\right)^{-2}
\end{equation}
where $r$ is the distance from the central source. Therefore, the area exposed to the unattenuated FUV strength $G_0$ is proportional to $r^2$, and thus, to $1/G_0$.

\begin{figure*}
	\includegraphics[width=1.0 \textwidth]{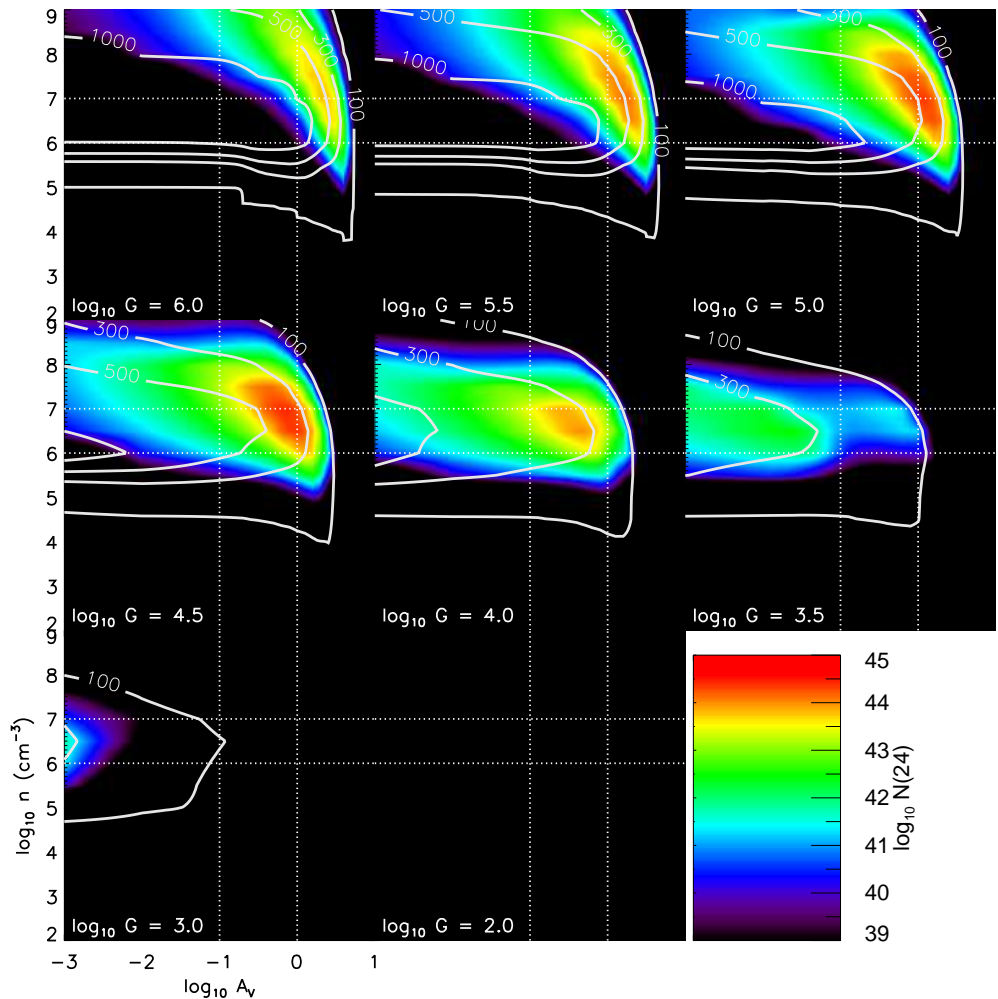}
	\caption{Rotational temperature $T_{\rm rot}$ (contour) and emitting CO number in $J$ = 24  $N(24)$ (image), as a function of  visual extinction ($A_{\rm V}$) and the total hydrogen density ($n$), for a given FUV strength in the Draine field.  $N(24)$ is calculated with the LVG model, and $T_{\rm rot}$ is fitted from $J$~=~14 to $J$~=~24 (see text). 
}\label{fig:draine_trot_n24}
\end{figure*}

\begin{figure*}
	\includegraphics[width=1.0 \textwidth]{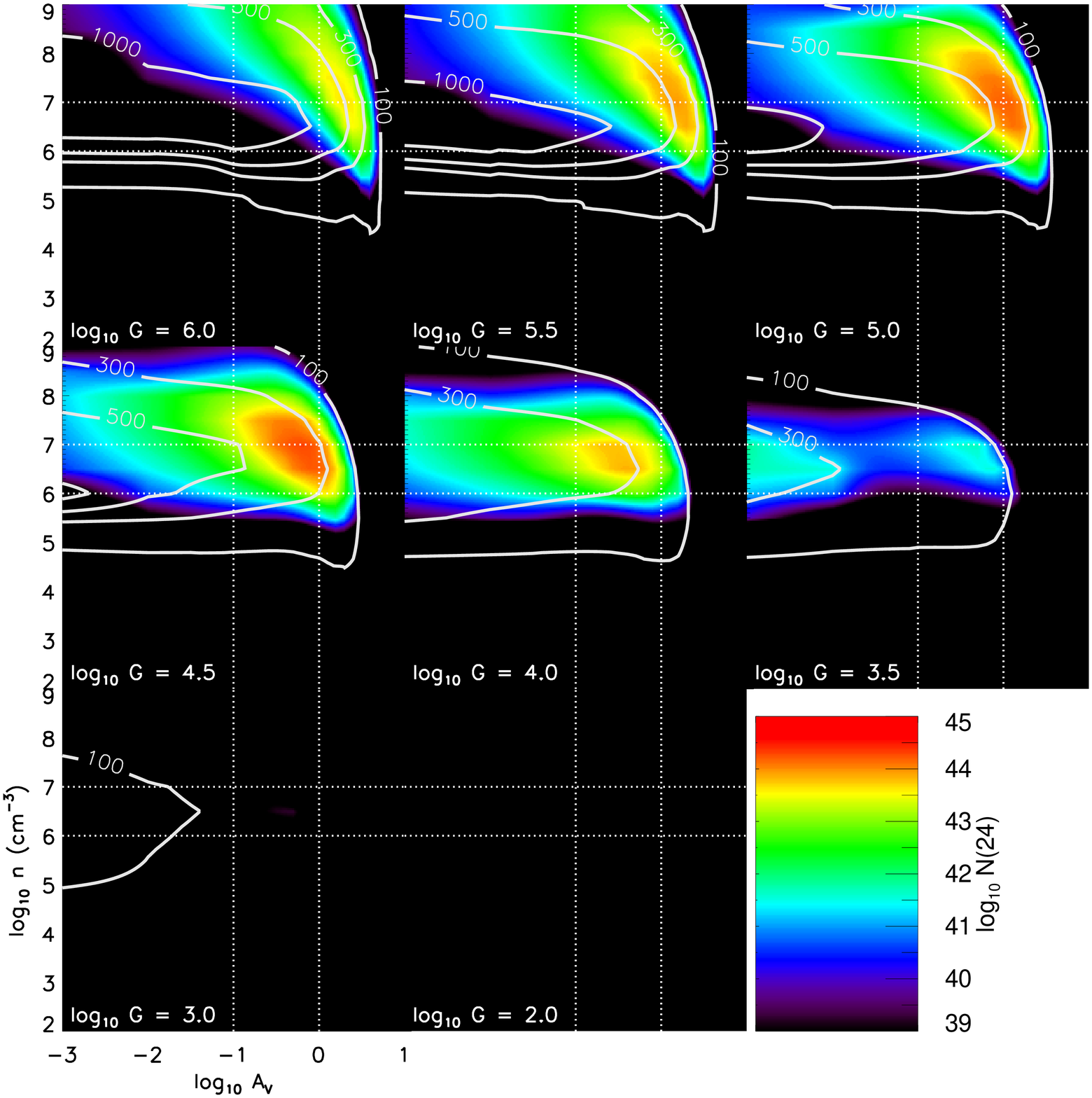}
	\caption{The same as Fig. \ref{fig:draine_trot_n24} except for BB1.5.}\label{fig:bb1.5_trot_n24}
\end{figure*}

\begin{figure*}
	\includegraphics[width=1.0 \textwidth]{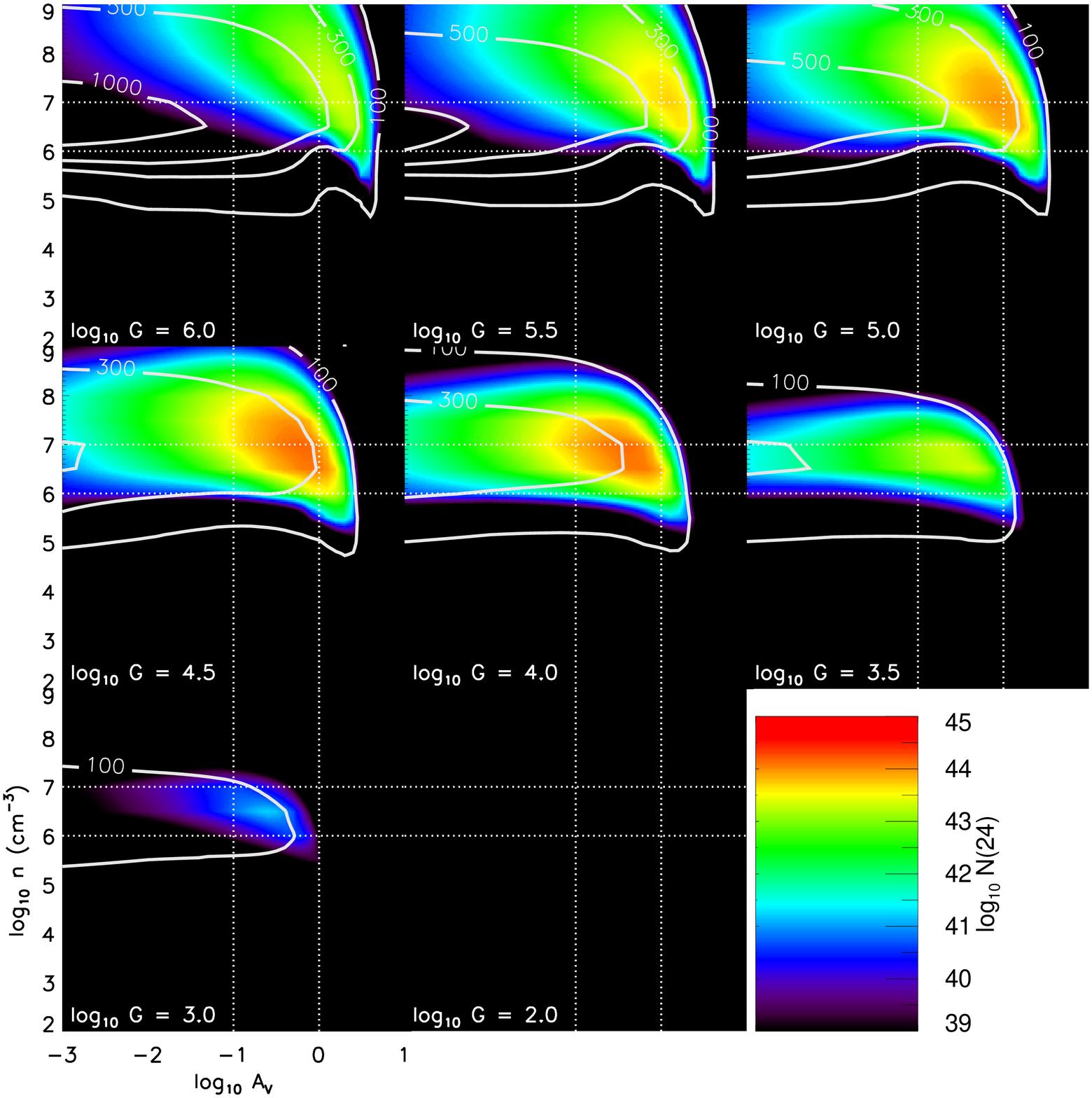}
	\caption{The same as Fig. \ref{fig:draine_trot_n24} except for BB1.0.}\label{fig:bb1.0_trot_n24}
\end{figure*}

 Figs.~\ref{fig:draine_trot_n24}~--~\ref{fig:bb1.0_trot_n24} show $N(24)$ and the rotational temperature $T_{\rm  rot}$ fitted from \J~=~14 to \J~=~24 for the models with the Draine field, BB1.5, and BB1.0, respectively. The CO $J = 24-23$ transition traces the warm component of $T_{\rm  rot} \ge 300$~K and is emitted from near the surface (0.1~$\le A_{\rm  V} \le$~1) of dense region (6~$\le \mathrm{log}\, n \le$~8) with high FUV strength (log~$G_0 \ge 3.5$). These regions are in a few hundred~AU from the protostar. When the FUV strength increases for the same density, for example, log~$n$~=~7, most fluxes of the mid-$J$ CO transitions are emitted with the similar $T_{\rm  rot}$ but from deeper $A_{\rm  V}$. This can explain why $T_{\rm  rot}$ has the universal value, independent of bolometric luminosity and density of embedded protostars.

\begin{figure*}
	\includegraphics[width=1.0 \textwidth]{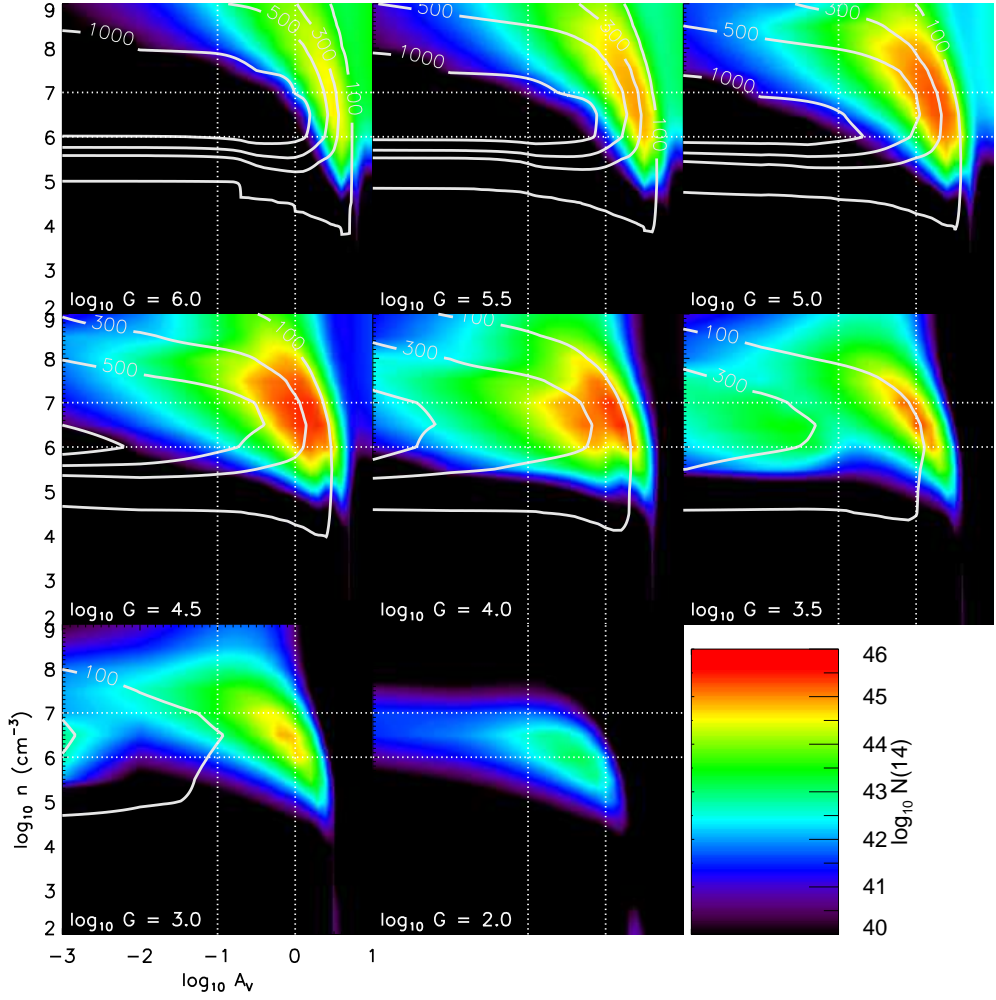}
	\caption{The same as Fig. \ref{fig:draine_trot_n24} except for N(14).}\label{fig:draine_trot_n14}
\end{figure*}

\begin{figure*}
	\includegraphics[width=1.0 \textwidth]{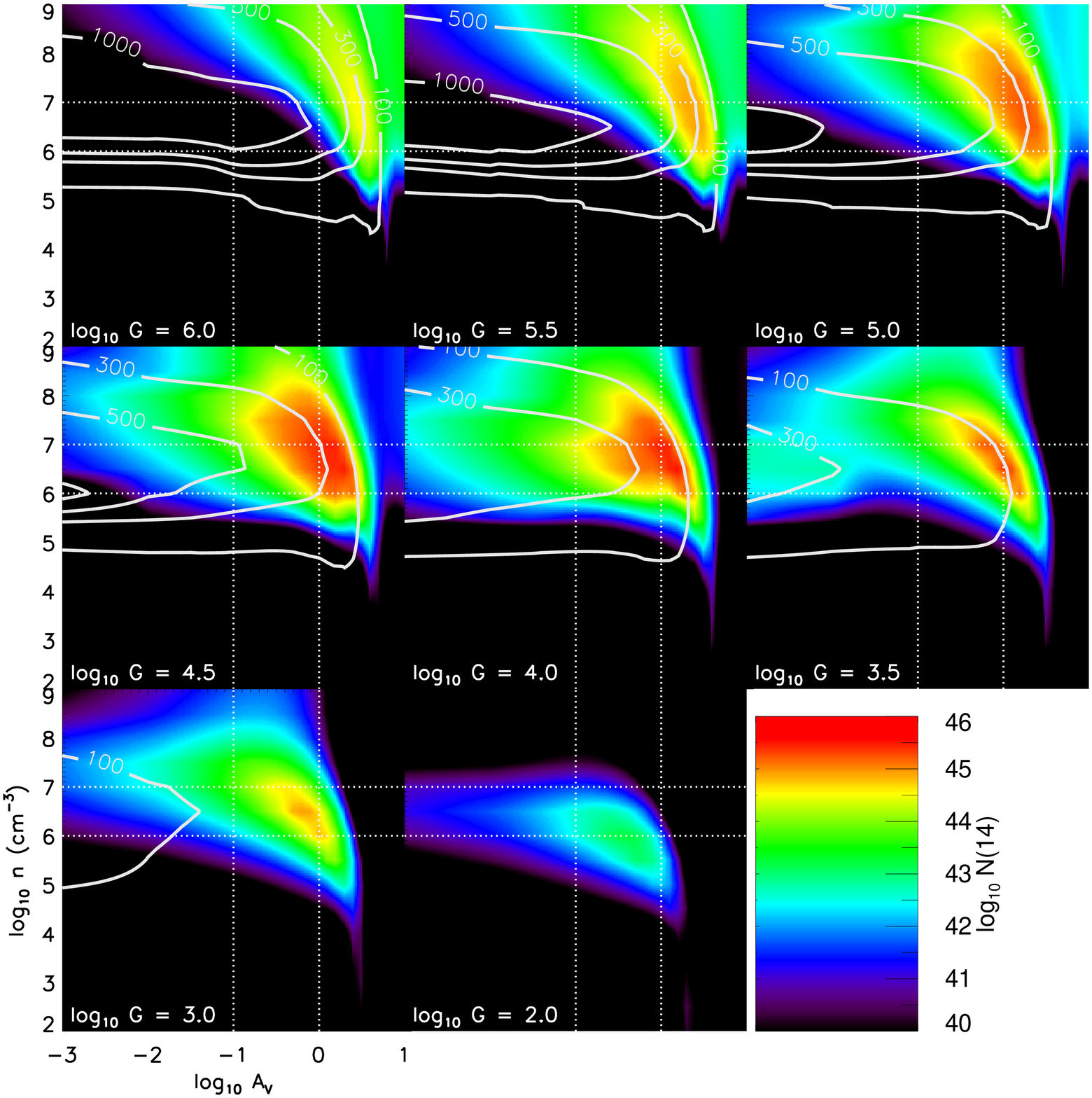}
	\caption{The same as Fig. \ref{fig:draine_trot_n14} except for BB1.5.}\label{fig:bb1.5_trot_n14}
\end{figure*}

\begin{figure*}
	\includegraphics[width=1.0 \textwidth]{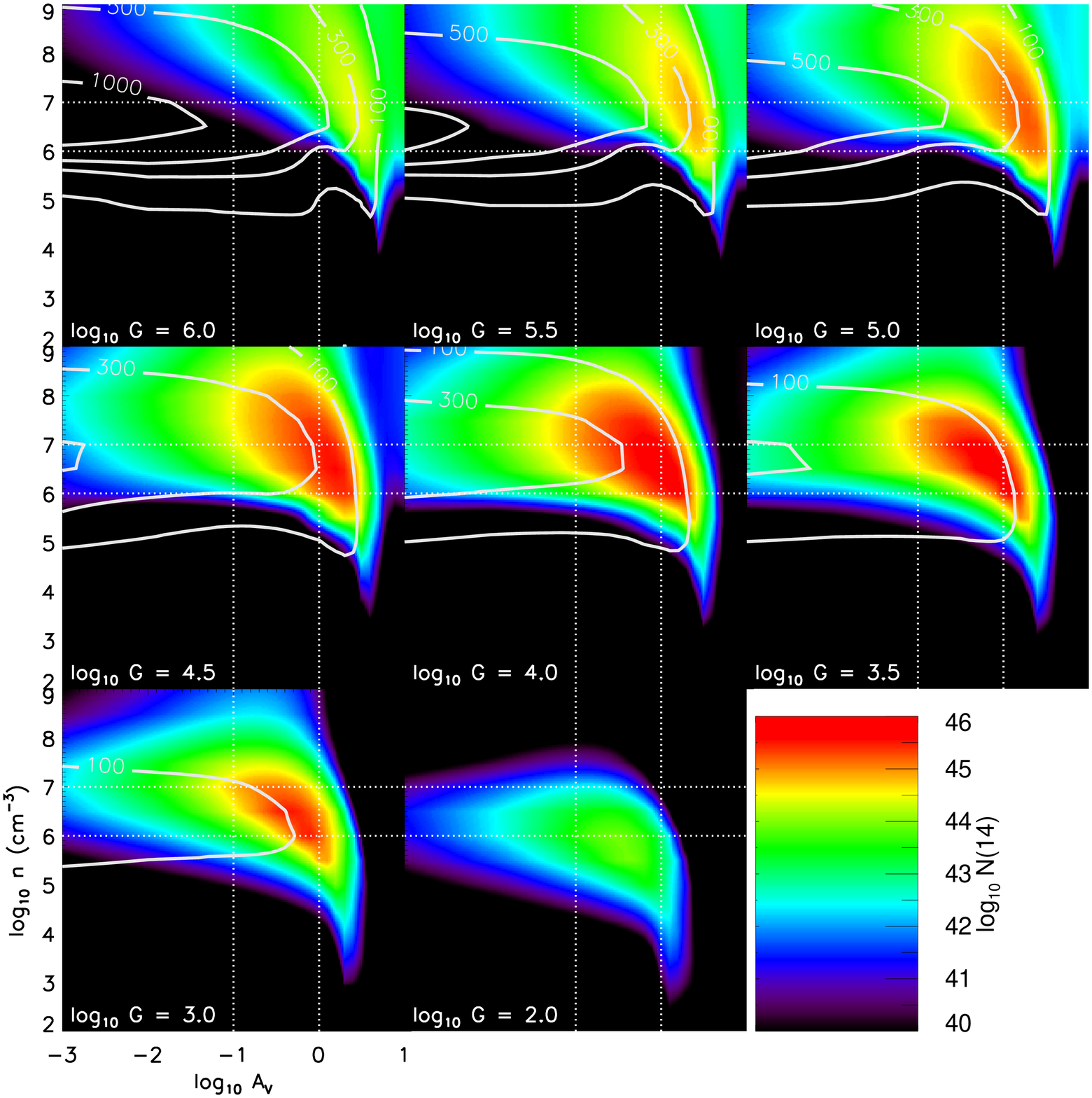}
	\caption{The same as Fig. \ref{fig:draine_trot_n14} except for BB1.0.}\label{fig:bb1.0_trot_n14}
\end{figure*}

 The CO $J = 14-13$ line is emitted from the deeper region than the CO $J = 24-23$ line (see Figs.~\ref{fig:draine_trot_n14} - \ref{fig:bb1.0_trot_n14}). As this line traces the cool component ($T_{\rm rot} \simeq 100$~K) as well as the warm one, we should run the 2D PDR models to check the PDR model can produce the FIR mid-$J$ CO lines observed by \Herschel/PACS. The models with BB1.5 and BB1.0 have higher $N$(24) and $N(14)$, but lower $T_{\rm  rot}$ than the model with the Draine field.  
 
\section{UV heated outflow cavity walls for HH46}\label{sec:application}

We have applied our PDR model to the UV-heated outflow cavity walls for HH46 following the models of \citet{Visser12} and \citet{bruderer2009}. The CO ladders observed by \Herschel/PACS in HH46 show that two temperature (warm and hot) gas components are indicative in the rotation diagram, and the warm component has $T_{\rm rot} \simeq 300$~K, which is possibly produced by UV photons \citep{Visser12}.

\subsection{Model}\label{sec:hh46model}
 A density distribution of the envelope is assumed to be a power law of the spherically symmetric one dimensional model, except for the outflow cavity. We adopt the density structure of envelope from \citet{Visser12},
  \begin{equation}
  n = 2.2 \times 10^9 \left(\frac{r}{34.6 \mathrm{AU}}\right)^{-2.0} \mathrm{cm}^{-3}.
  \end{equation}
 The outflow cavity is carved out with the opening angle of 60$^\circ$ by Eq. \ref{eq:bound}. We assume that the density inside the outflow cavity is $1.2\times 10^{4}$~cm$^{-3}$ \citep{Neufeld2009,Visser12}.

\begin{figure*}
\includegraphics[width=0.495 \textwidth]{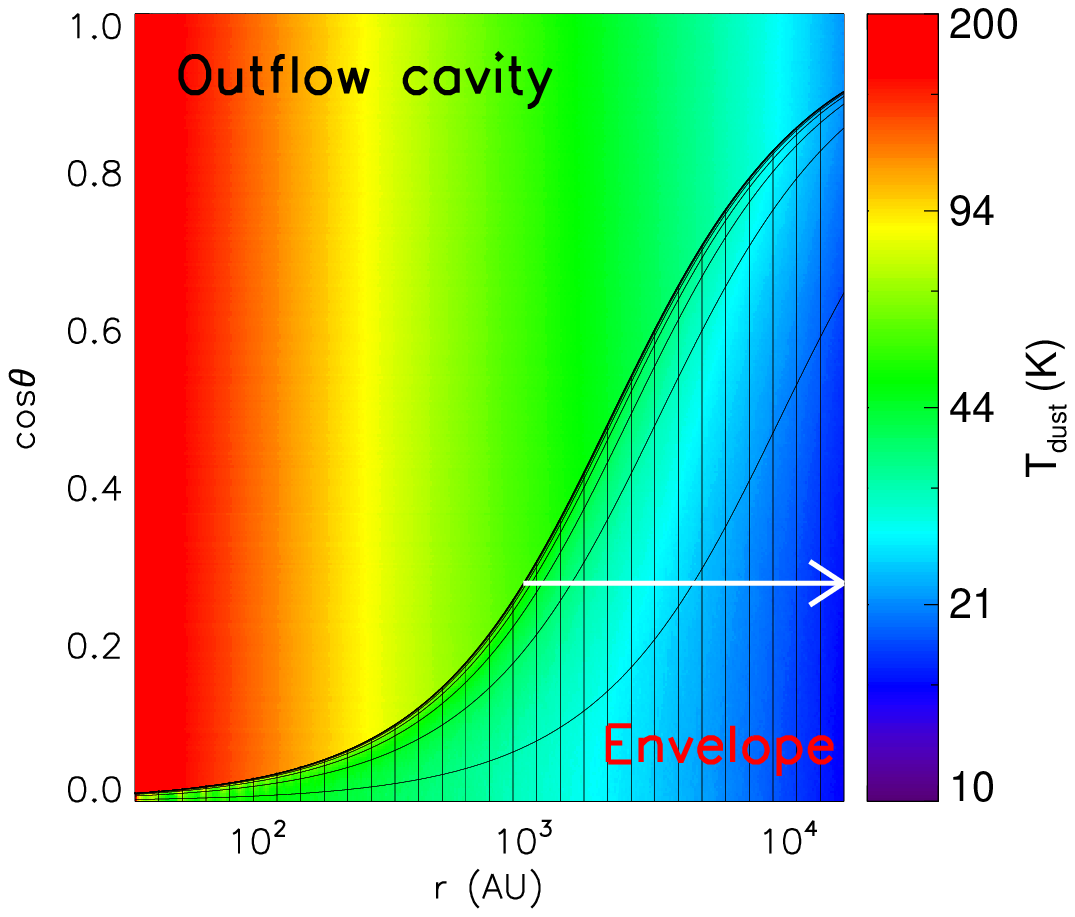}
\includegraphics[width=0.495 \textwidth]{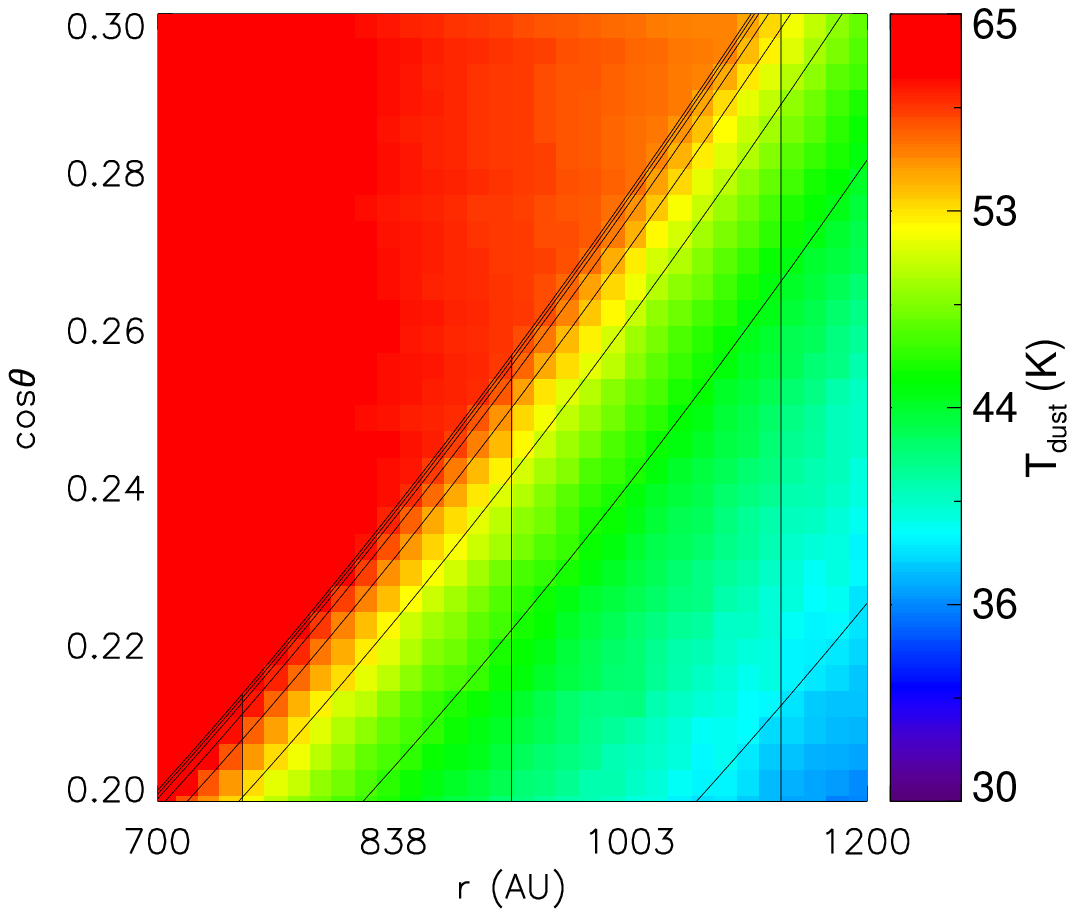}
\includegraphics[width=0.495 \textwidth]{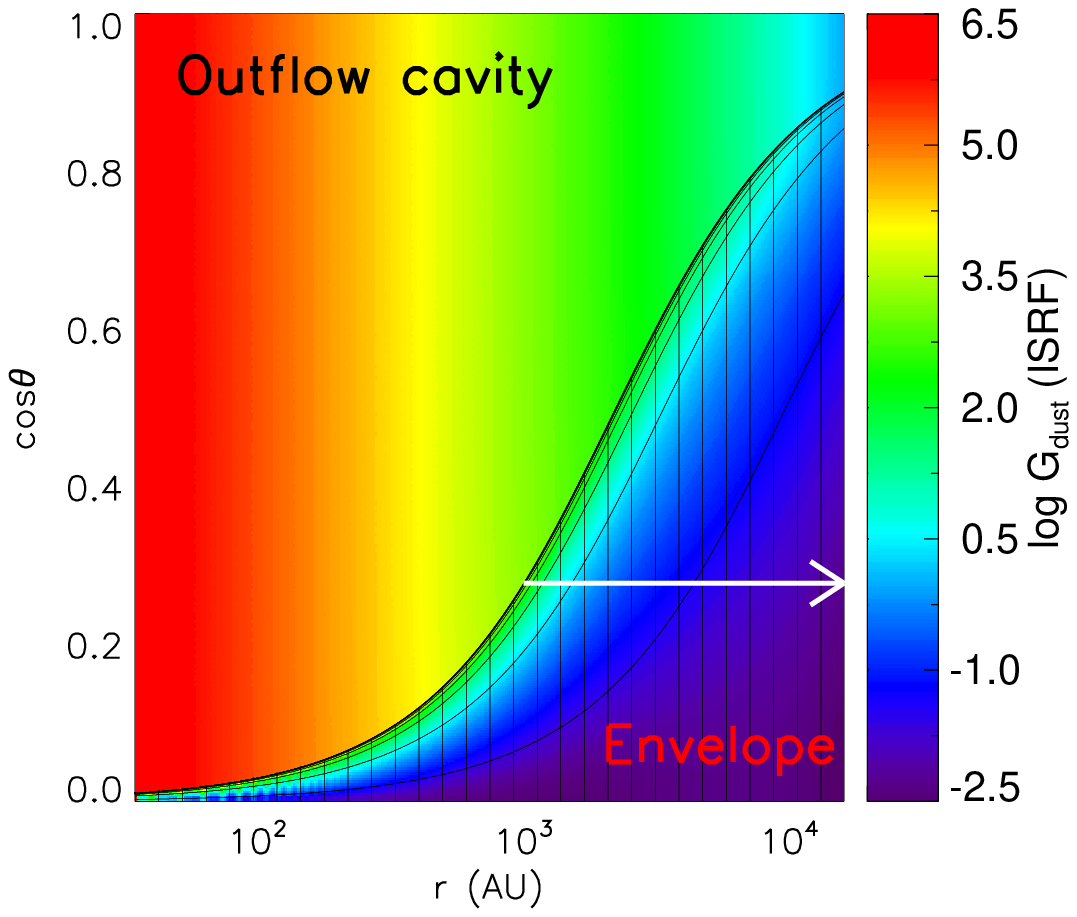}
\includegraphics[width=0.495 \textwidth]{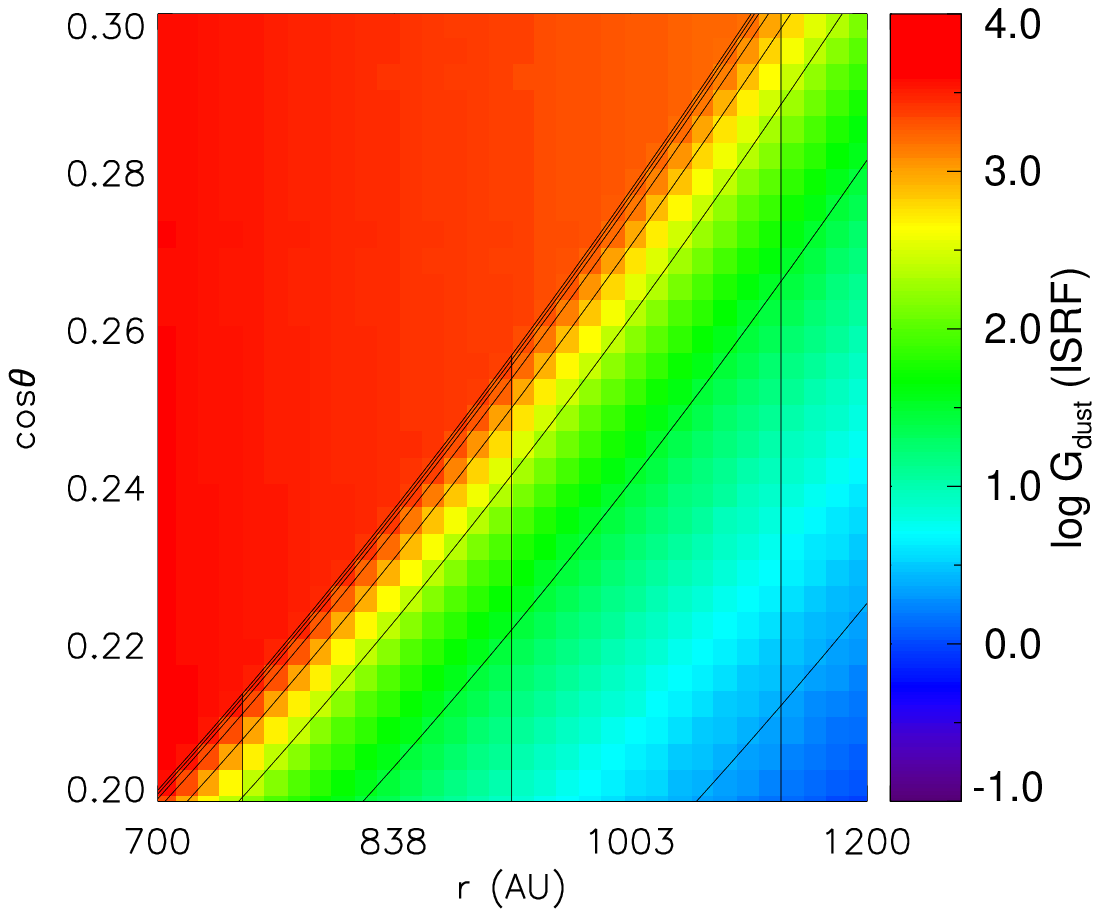}
\caption { The r and cos$\theta$ diagrams for the dust temperature $T_{\rm dust}$ and the attenuated FUV strength $G_{\rm dust}$ calculated with RADMC. The $r$ and $\delta$ grids are plotted as vertical and curved lines, respectively. Right columns zoom in to the sub-region near the surface of the inner envelope. The $\delta$ grids describe well the distributions of dust temperature and the attenuated FUV strength in the deep envelope (left) as well as narrow outflow cavity wall surfaces (right) for the envelope model with outflow cavity. A column density is measured along the horizontal white arrow for setting the $\delta$ grids (see Sec.~\ref{sec:hh46model}). } \label{fig:radmc}
\end{figure*}

The dust temperature is calculated with RADMC-3D adopting the same dust opacity used in Sec.~\ref{sec:fuvrt}. We choose the stellar temperature of 5000~K, which does not significantly affect the dust temperature in the envelope \citep{Visser12}. The bolometric luminosity of 27.9~L$_{\odot}$ is adopted as the luminosity of the internal source \citep{karska13}. Fig.~\ref{fig:radmc} (upper panels) shows the dust temperature distribution  calculated by RADMC-3D in the 2D spherical coordinate system ($r$, $\theta$) with 360 (in $r$) $\times$ 300 (in $\theta$) grid cells. 
For the PDR model, we use the ($r$, $\delta$) coordinates, where $r$ is the radial distance from the central protostar and $\delta$ is defined in Sec.~\ref{sec:raytrace}. 
 The dust temperature $T_{\rm dust}(r, \theta)$ from RADMC-3D is, therefore, converted to $T_{\rm dust}$ in the ($r$,~$\delta$) coordinates by
\begin{equation}
T_{\rm dust} = \frac{\int_{r_{min}}^{r_{max}} r^2 \int_{cos \theta (r,\delta_{min})}^{cos \theta (r,\delta_{max})} T_{\rm dust}(r,\theta)n(r,\theta) d(cos\theta)}{\int_{r_{min}}^{r_{max}} r^2 \int_{cos \theta (r,\delta_{min})}^{cos \theta (r,\delta_{max})}n(r,\theta) d(cos\theta)},
\end{equation}
where $n(r,\theta)$ is the total hydrogen number density and $cos \theta (r,\delta)$ is the $cos \theta$ for the given $(r,\delta)$ grid point.
The $r$ and $\delta$ grids are plotted as vertical and curved lines, respectively, in Fig.~\ref{fig:radmc}. The dust temperature distribution is well described by the ($r$, $\delta$) grids.
  
 The $\delta$ coordinate is an adequate coordinate system to describe the outflow structure and resolve the very narrow regions near the outflow wall surface where the warm CO gas exists as shown in Fig.~\ref{fig:radmc}. In addition both PDR and non-LTE line radiative transfer calculations should be able to deal with scales ranging from $\sim10$~AU to $\sim10^4$~AU. 
  Properties of the PDRs are characterized by three parameters: a density, an unattenuated (or incident) FUV strength, and  a depth ($A_{\rm V}$ or a column density of hydrogen). As the density profile of the envelope is assumed to be a power-law of radial distance and the incident FUV strength follows an inverse square law of the radial distance, the radial distance $r$, which is equidistant  on logarithmic scale as shown as vertical lines in Fig.~\ref{fig:radmc}, has been chosen.  
 
 $A_{\rm V}$ is described by a column density in the 1D models. As photons from the central protostar propagate radially (horizontally in Fig.~\ref{fig:radmc}), radial points are first calculated to have an equivalent interval in log $A_{\rm V}$ from the outflow wall surface  along the horizontal direction at $r$=1000 AU as shown with the white arrow in Fig.~\ref{fig:radmc}, and then, the equation~(\ref{eq:Av}) is used to trace $\delta$ grids for the found radial points  and cos $\theta$ of 0.28.
At $r$=1000 AU, the $\delta$ point closest to the outflow surface has $A_{\rm V}$=0.1, and the deepest $\delta$ point has $A_{\rm V}$=10 in Fig.~\ref{fig:radmc}.  
 
 The $\delta$ coordinate presents well the very narrow layer near the outflow surface for $A_{\rm V} \le 1$, where the mid-$J$ CO emission is radiated as seen in the 1D models. 
 In addition, the FUV strength near the surface drops by an order of magnitude (see the bottom right panel of Fig.~\ref{fig:radmc}), requiring a high resolution. Therefore, we calculate the FUV radiative transfer in the ($r$, $\delta$) coordinate system instead of using the result of RADMC-3D 
  because 30 (in log~$r$)$\times$ 10 (in $\delta$) grid cells can provide a higher spatial resolution (for the FUV strength near the surface) than 360 (in log $r$) $\times$ 300 (in cos$\theta$)  in the spherical coordinates.   
 However, the dust temperature is calculated by RADMC-3D 
 because the dust temperature varies only within 10 K near the surface, and a higher resolution does not affect results.

 As the mid-$J$ CO emission is the major concern for this model, half of the $\delta$ grids are put near the surface to provide a enough spatial resolution in the UV heated cavity walls.
  A larger number of grid (90 in log~$r$ and 30 in $\delta$) shows a similar result to the 300 grid model. We note that our PDR model with  300 grids cannot describe sharp transitions of H-H$_2$ and C$^+$-C-CO, which do not affect our result.
     
  We assume that the central protostar is the only FUV source. The initial FUV radiation field stored in each photon package is given by,
\begin{equation}
I_0=\frac{L_{\rm  uv} }{N_{\rm  phot}}
\end{equation}
where $L_{\rm  uv}$ is the FUV luminosity of the central protostar and $N_{\rm phot}$ is the number of photons. The photon packages initially propagate the system in the radial direction, and they are traced until escaping from the outer boundaries of both the outflow cavity and the envelope.  

\begin{figure}
\includegraphics[width=0.9 \textwidth]{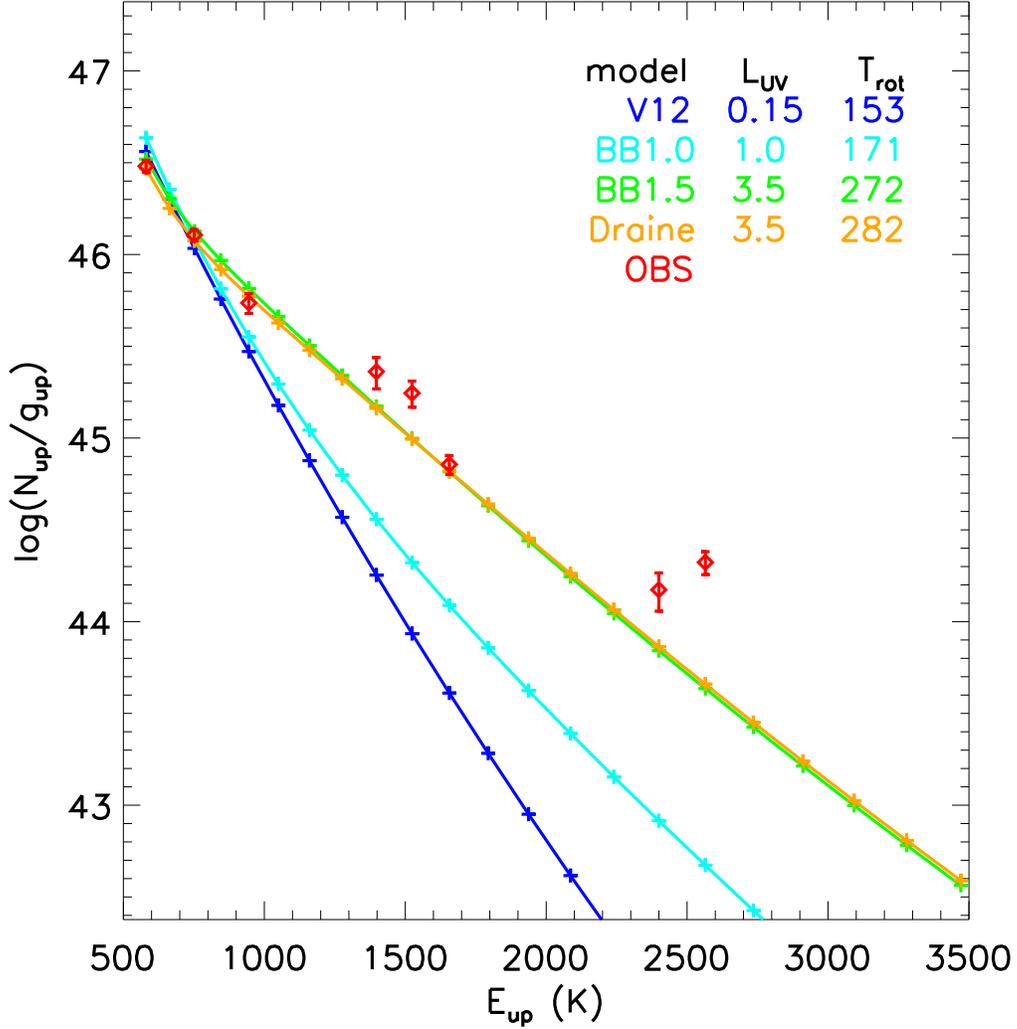}
\caption{ Rotational diagram of models for Each UV radiation field. The blue line indicates the model calculated with the same method as \citet{Visser12}. The cyan, green, and orange lines indicate the model with BB1.0, BB1.5, and Draine field, respectively (see text). \Herschel/PACS observation data are plotted as the red diamonds. Their rotational temperatures are fitted up to $E_{\rm up} \le 1,800$~K, and the best fit UV luminosities in units of $L_{\rm UV}^{Y}$(= 0.7 L$_{\odot}$) and rotational temperatures are presented inside the box.}\label{fig:hh46_radiation}

\end{figure}

 As presented in Fig. \ref{fig:hh46_radiation}, we have run a comparison model as well as our self-consistent models for the different UV radiation fields (BB1.0, BB1.5, and the Draine field).  For the comparison model (Fig.~\ref{fig:hh46_radiation}), we have followed the method of \citet{Visser12} (hereafter V12 model). In this method, the gas temperature has been calculated from an approximated formula, $T(G_{\rm 0}, A_{\rm  V})~=~T_{\rm S}~\mathrm{exp}(-0.6~A_{\rm  V})$, where the surface temperature $T_{\rm S}$ was adopted from \citet{kaufman99}, and the chemistry has been calculated with BB1.0. 

  The FUV observation toward classical T Tauri stars shows that  the UV luminosity integrated from 1250 \AA\,  to 1750 \AA\, ($L_{\rm  uv}^{\rm Int}$) is related with the accretion luminosity ($L_{\rm  acc}$) as ${\rm log} {L}_{\rm  UV}^{\rm Int} =   0.836\times {\rm log}{L}_{\rm  acc} \,-\,1.67$ with an accuracy of 0.38 dex  \citep{yang12}. As the FUV luminosity integrated from 912 \AA\, to 2050 \AA\, is about 2 times of $L_{\rm  uv}^{\rm Int}$ for TW~Hya and AU~Mic \citep{yang12} and the accretion luminosity dominates the bolometric luminosity during the class 0 and I, we adopt a reference UV luminosity of $L_{\rm  UV}^{\rm Y} = 0.7\, \mathrm{L}_{\odot}$ (0.02~$L_{\rm bol}$).
  
  Level population and spectral images are calculated with RIG, which can solve the problem with the same coordinates used in the PDR model. The CO molecular data file \citep{Neufeld2012} is adopted from the Leiden Atomic and Molecular Database\footnote{http://home.strw.leidenuniv.nl/$^\sim$moldata/datafiles/co@neufeld.dat} \citep{Scho05}.
  Spectral images at the known inclination of HH46 ($i$=53$^\circ$) are synthesized with a spatial resolution of 0.$''$05 (23 AU at distance of HH46) and a spectral resolution of 0.1 km~s$^{-1}$. As most emission is from near the center of image, the intensity over the PACS 5$\times$ 5 spaxels (50$'' \times$50$''$) is summed. 

\subsection{Results}
In this section, we find the best fit UV luminosity inferred from our models that fit the \Herschel/PACS observations. The rotational diagrams from CO ladders detectable with \Herschel/PACS are plotted in Fig. \ref{fig:hh46_radiation}. The number of CO emitting in the $J$ level is calculated as following \citet{Green2013}:
\begin{equation}
\label{eq:NJ}
\mathcal{N}_{\rm OBS}(J)=\frac{4\pi D^{2}F_{J}}{h\nu_J A},
\end{equation}
where $F_{J}$ and $\nu_J$ denote the line flux and the frequency of the CO rotational transition from $J$ to $J-$1, $D$ is the distance to the source, $A$ is the Einstein coefficient, and $h$ is Planck's constant. 

\begin{figure*}
\includegraphics[width=0.495 \textwidth]{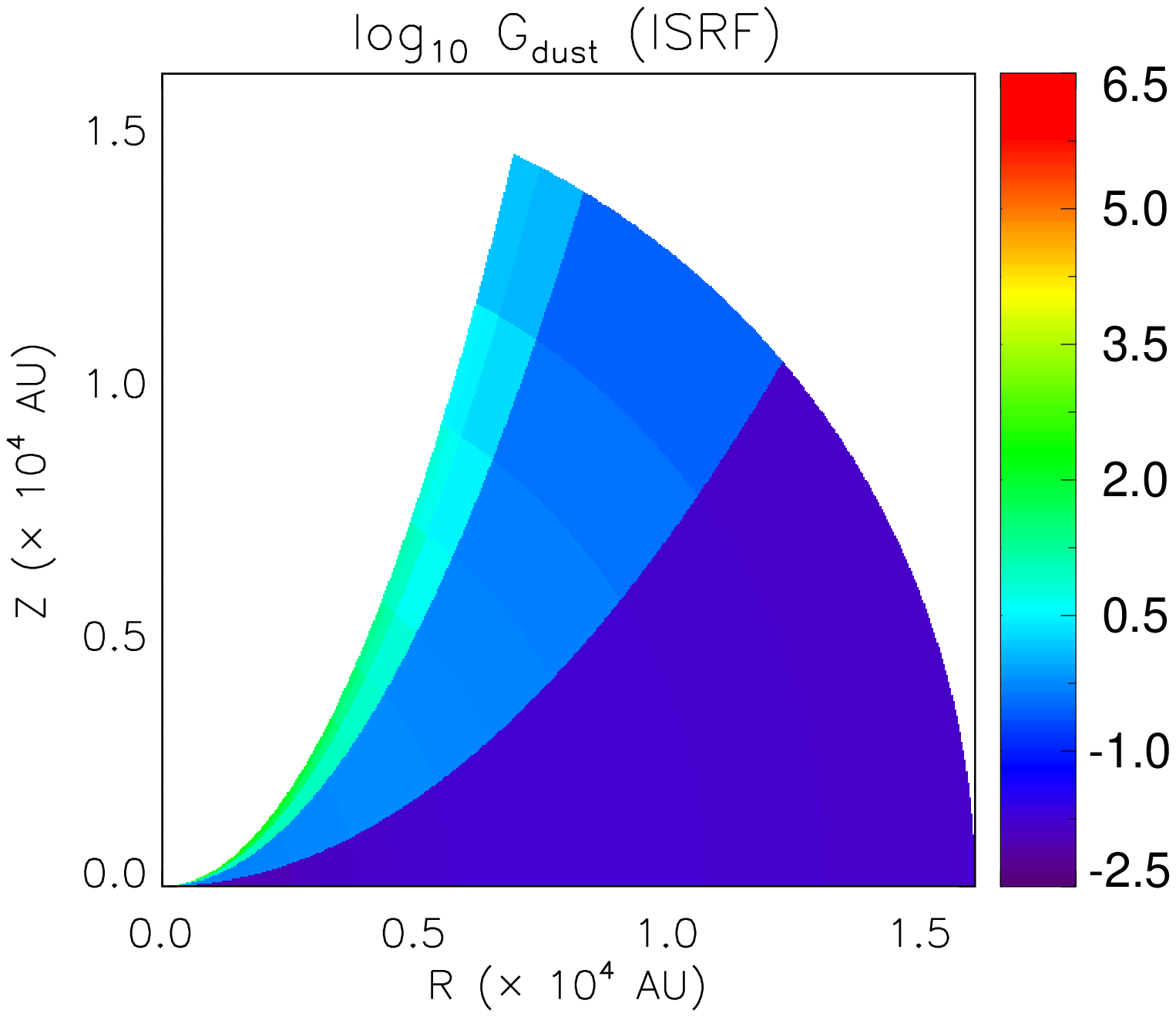}
\includegraphics[width=0.495 \textwidth]{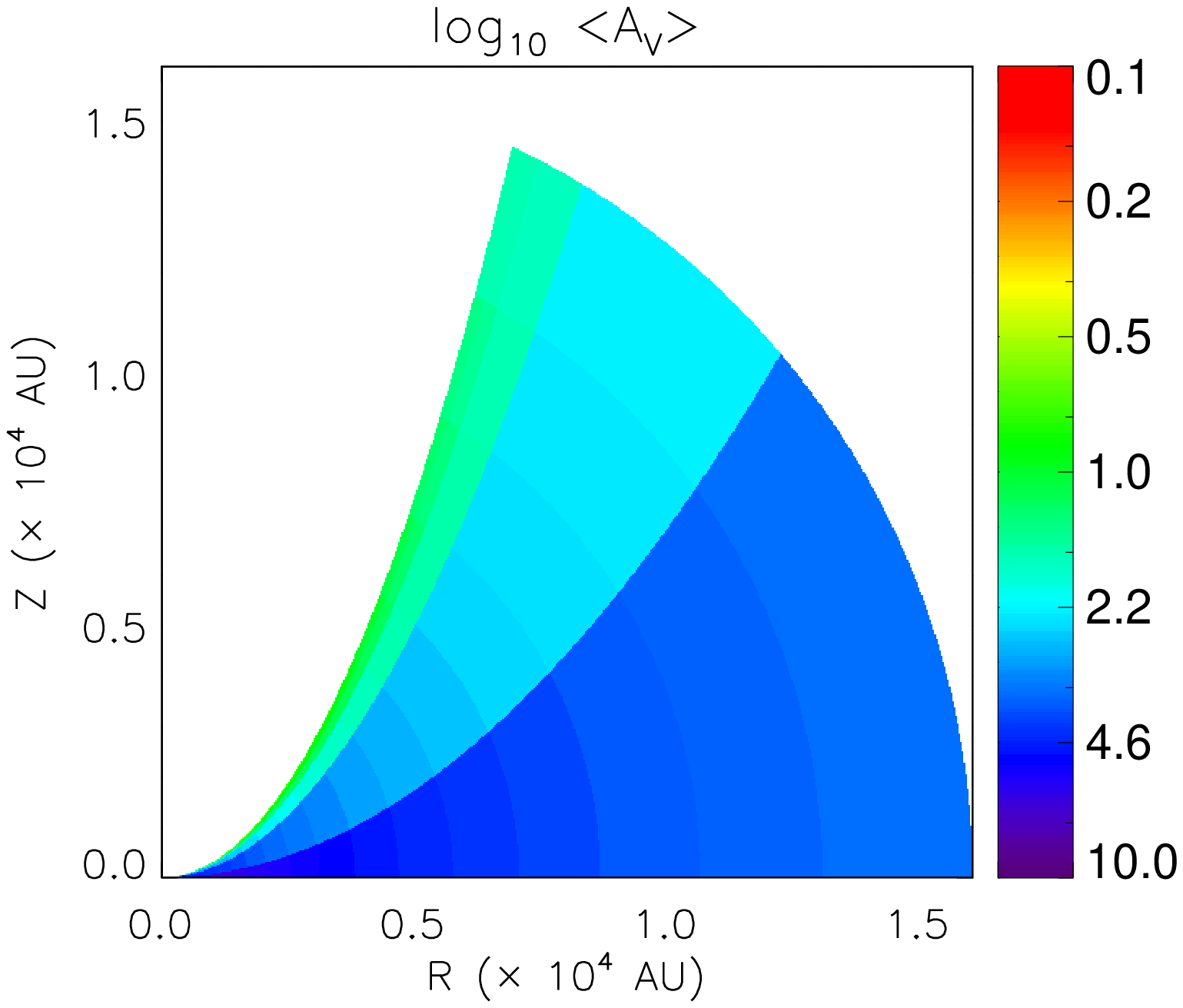}
\includegraphics[width=0.495 \textwidth]{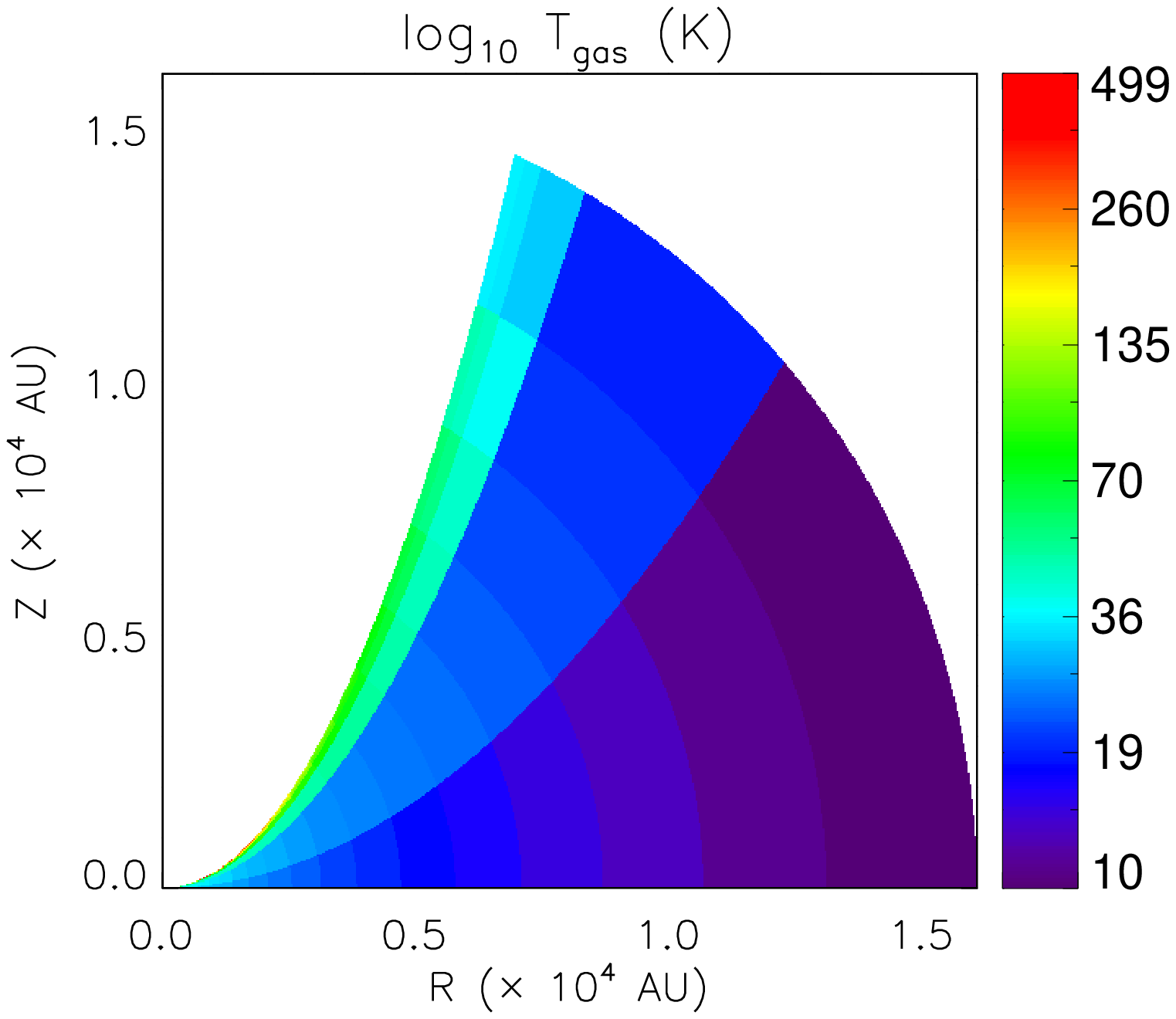}
\includegraphics[width=0.495 \textwidth]{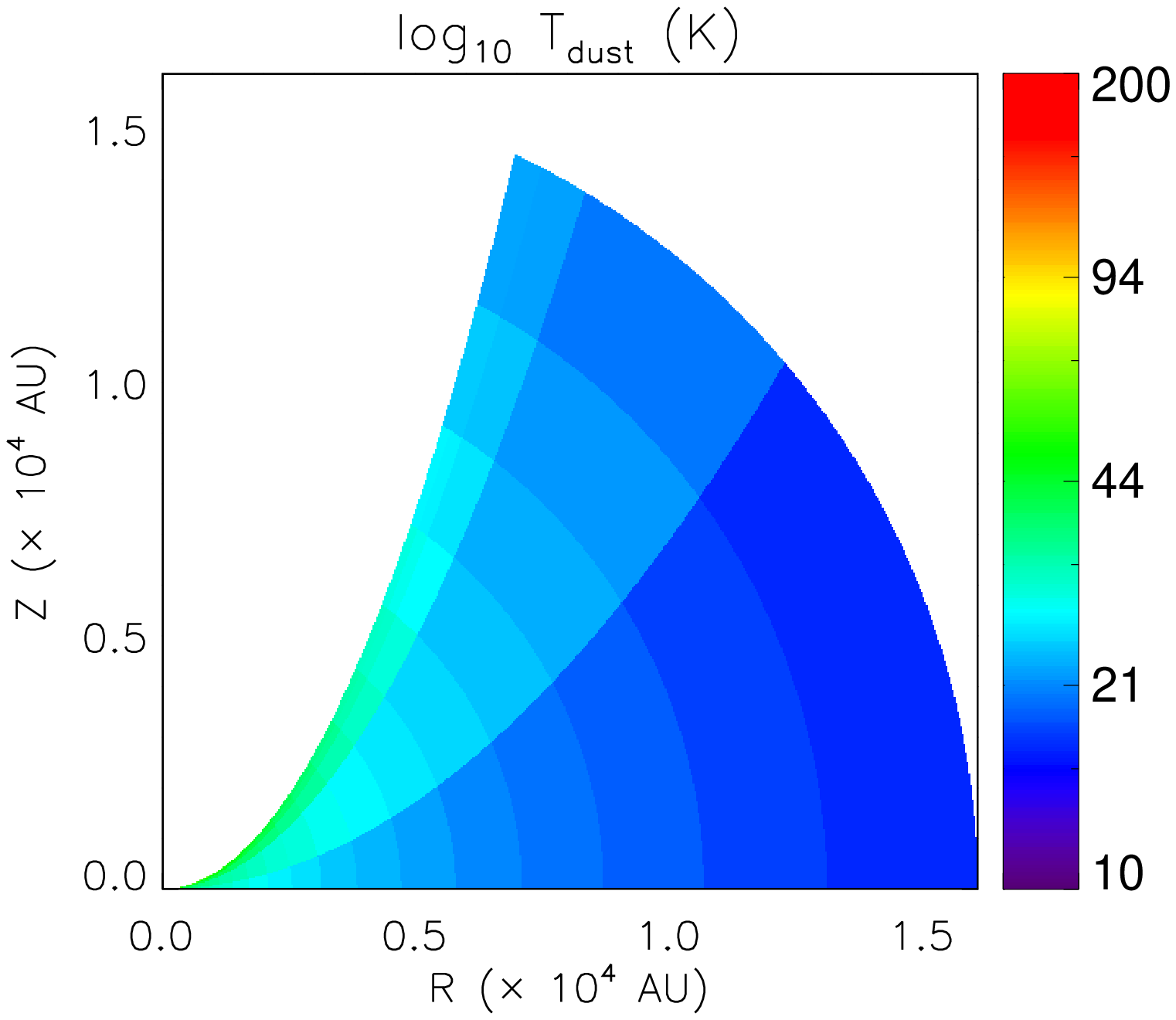}
\caption {Physical structure of HH46 with BB1.5: dust attenuated FUV strength (top left), average visual extinction $\left<A_\mathrm{V}\right>$ (top right), gas temperature (bottom left), and the dust temperature structure (bottom right).
The gas temperature is calculated with our PDR model (Sec.~\ref{sec:model}), but the dust temperature is calculated with RADMC-3D (Sec.~\ref{sec:hh46model}).
} \label{fig:hh46_2d}
\end{figure*}

 Fig.~\ref{fig:hh46_2d} shows 2D structure of the dust attenuated FUV strength $G_{\rm dust}$ (top left), average visual extinction $\left<A_{\rm V}\right>$ (top right), gas temperature $T_{\rm gas}$ (bottom left), and dust temperature $T_{\rm dust}$ (bottom right)  for the HH46 model. 
 The color scales of $G_{\rm dust}$ and $T_{\rm dust}$ is the same as those in Fig.~\ref{fig:radmc}, and $G_{\rm dust}$ calculated by RADMC-3D and our code are similar to each other. 
 $\left<A_{\rm V}\right>$ along the equatorial plane decreases outward because $G_{\rm dust}$ is almost constant in this scale while unattenuated FUV strength $G_0$, which can be approximated by the dust attenuated FUV strength in the outflow cavity, drops because of the inverse square  law of distance (see Eq.~\ref{eq:Av}).
 Because scattered UV photons that come through the surface in higher z pass lower column densities 
 compared to photons passing through near the equatorial plane. As a result, $G_{\rm dust}$ is nearly constant in large scale. 
 
\begin{figure*}
\includegraphics[width=1.0 \textwidth]{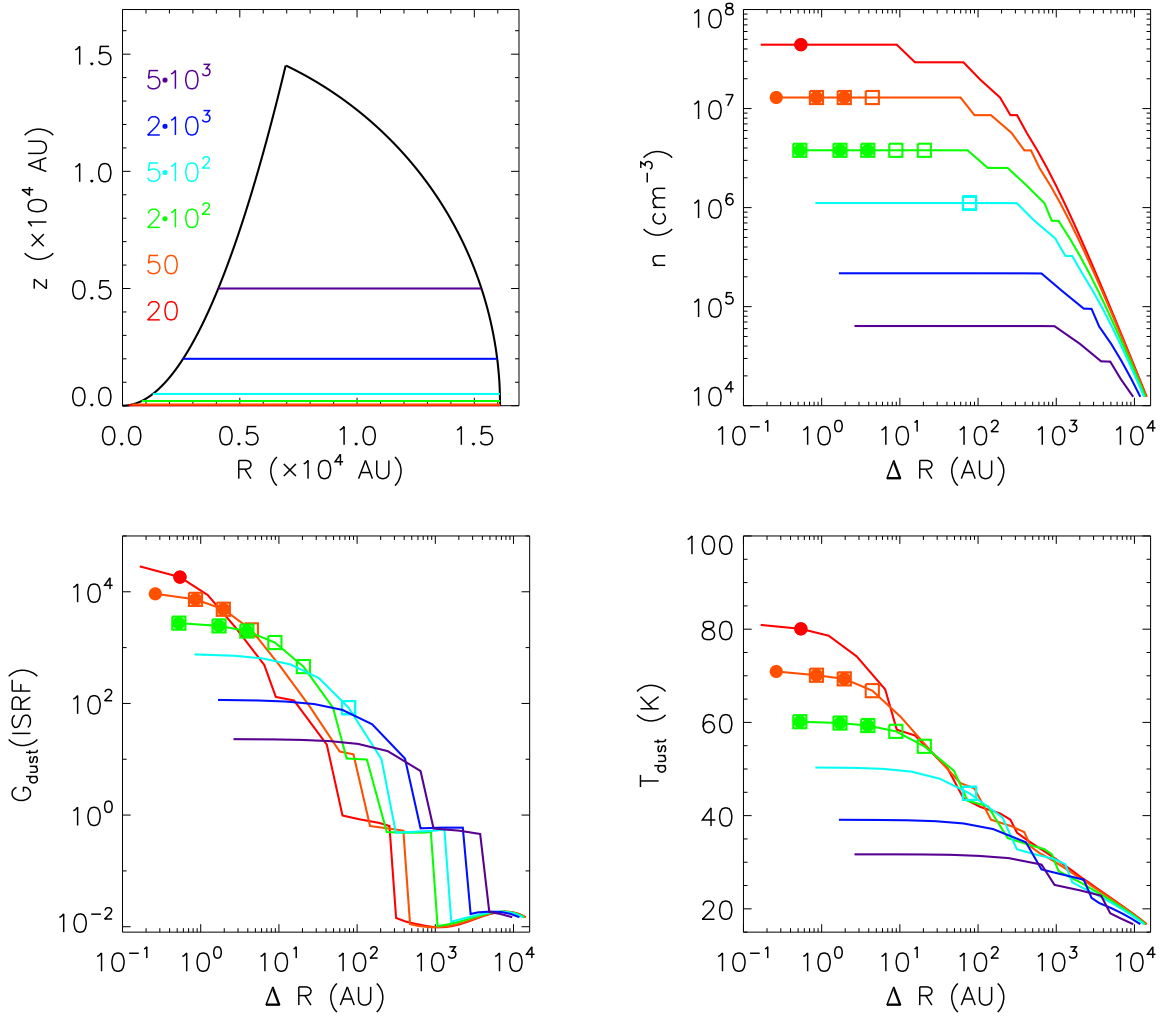}
\caption { Density(top right), FUV strength (bottom left), and dust temperature (bottom right) distributions along given horizontal cuts in the envelope of HH46. Each color line indicates the physical values for a given z-height, which is represented with the same color in the top left panel. $\Delta R$ is the horizontal distance from the outflow cavity wall surface. The filled circles, open squares, and filled squares on top of the lines indicate the grid cells where emissions of $J$=24--23, 14--13, and both lines are radiated, respectively. } \label{fig:hh46_grid}
\end{figure*}

\begin{figure*}
\includegraphics[width=1.0 \textwidth]{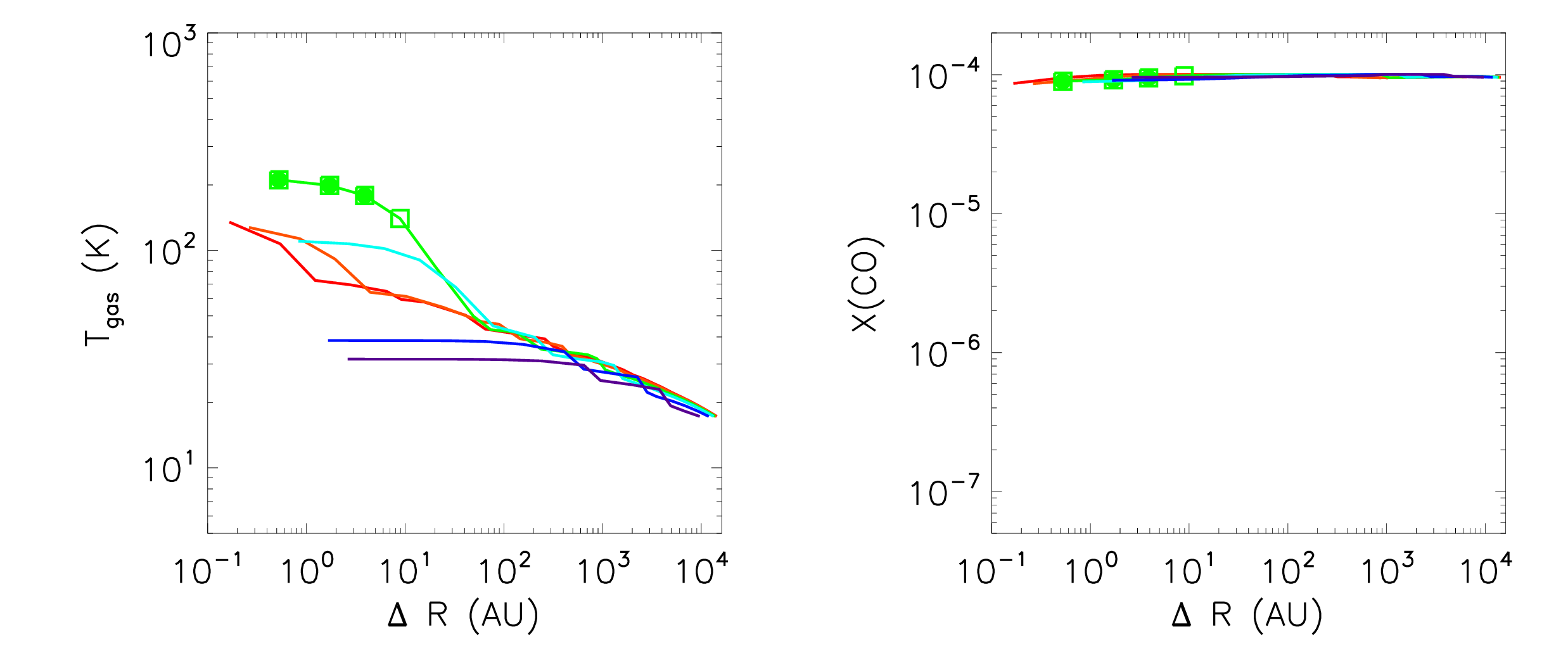}
\includegraphics[width=1.0 \textwidth]{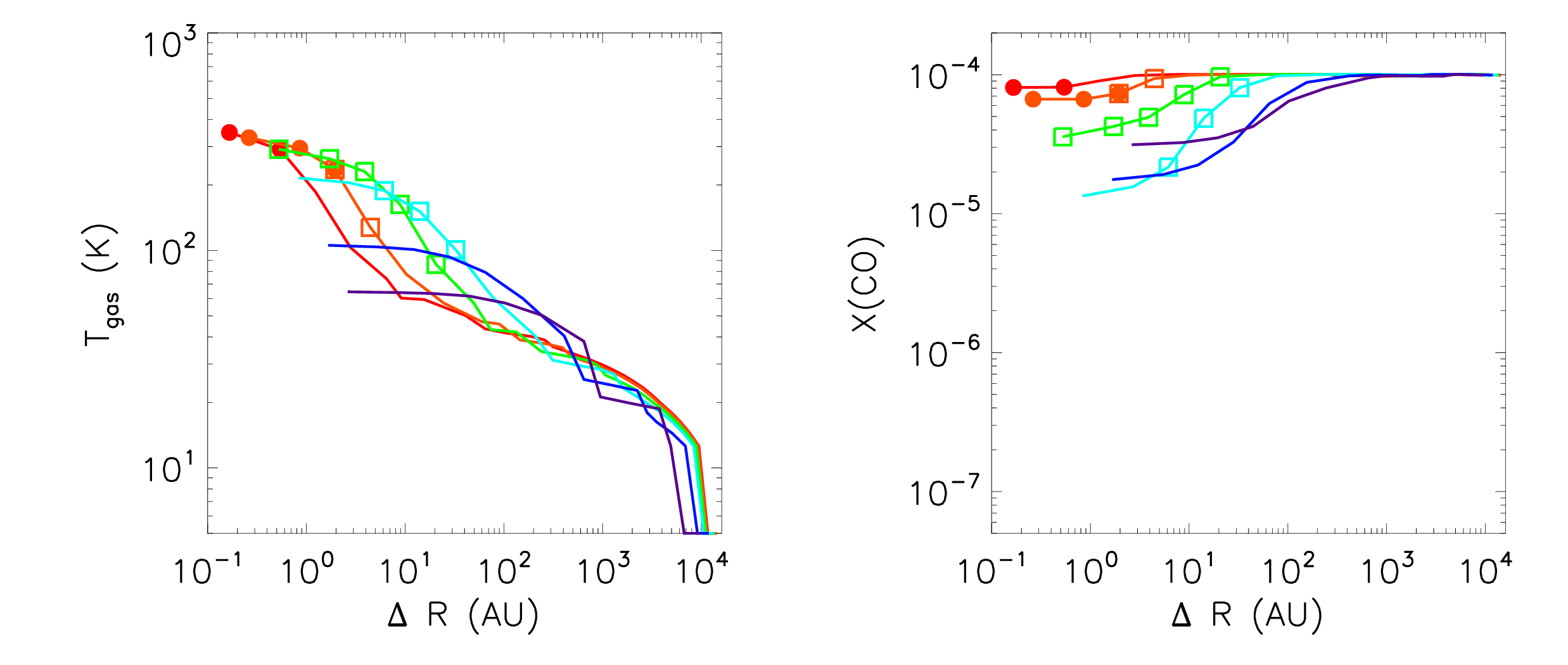}
\caption {Gas temperature (left column) and CO abundance (right column) of best-fit models. Top and bottom panels indicate the model for V12   and  BB1.0, respectively. Color lines are the same as presented in Fig. \ref{fig:hh46_grid}} \label{fig:hh46_result1}
\end{figure*}

\begin{figure*}
\includegraphics[width=1.0 \textwidth]{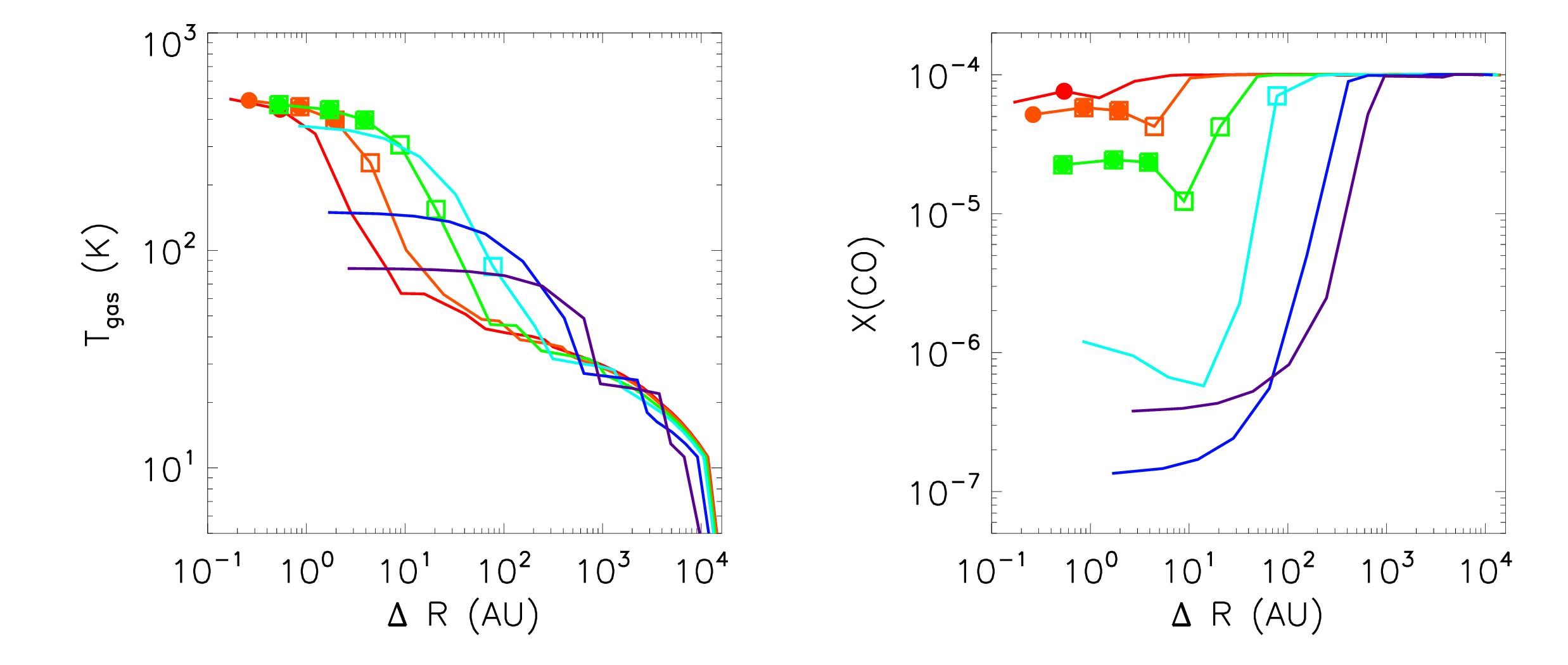}
\includegraphics[width=1.0 \textwidth]{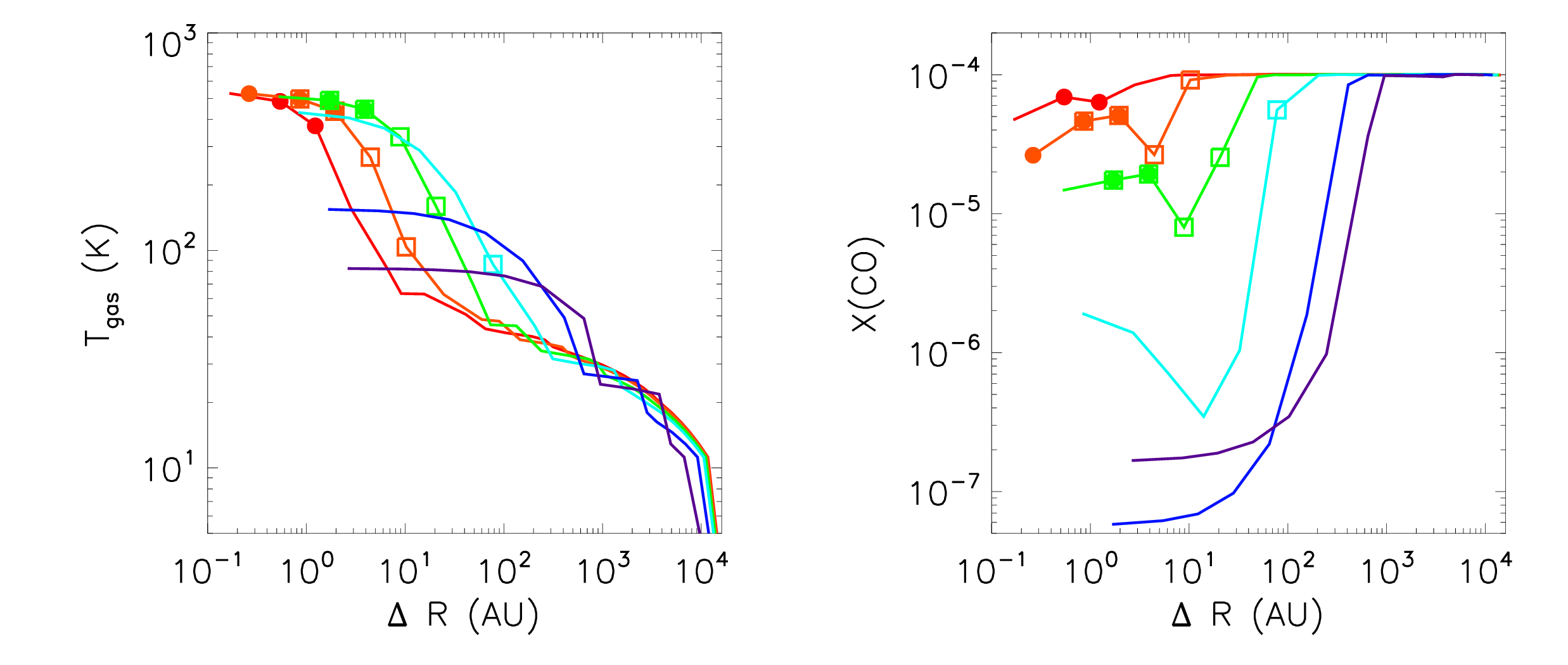}
\caption {The same as Fig. \ref{fig:hh46_result1} except for the best fit models of BB1.5 (top) and the Draine field (bottom).} \label{fig:hh46_result2}
\end{figure*}

 Horizontal distributions of physical and chemical properties for given z-heights are plotted in Figs.~\ref{fig:hh46_grid}-\ref{fig:hh46_result2}. 
 The given z-heights are marked with horizontal color lines and the same color texts
 in top left panel of Fig.~\ref{fig:hh46_grid}. 
 Other panels show the physical values along the horizontal lines from the surface of the outflow cavity wall for given z-heights. Filled circles and open squares plotted over each line  indicate grid cells where most of emissions of $J$=24--23 and $J$=14--13 are radiated, respectively. The filled squares in the plots present the grid cells where both lines of $J$=24--23 and $J$=14--13 contribute to the total emission similarly.
   $J$~=~14 and 24 are the lowest and highest upper levels for the representative transitions in the warm component of CO gas. 
  If a grid has a volume of $V$, the CO abundance of X(CO), and the population in the $J$ level $n(J)$, the normalized number of CO in the $J$ level $N^n(J)$ is defined as follows:
\begin{equation}
N^n(J) =\frac{n(J)\, \mathrm{X(CO)}\, V}{ \mathcal{N}_{\rm OBS}(J)}
\end{equation}
where $\mathcal{N}_{\rm OBS}(J)$  is the observed value described in Eq.~\ref{eq:NJ}.
We note that the distribution of the density (top right of Fig.~\ref{fig:hh46_grid}), the FUV strength (bottom left of Fig.~\ref{fig:hh46_grid}), and the dust temperature (bottom right of Fig.~\ref{fig:hh46_grid}) are the same for all models of different UV spectral types.
However, the absolute values of $G_{\rm dust}$ vary with the UV luminosity that fits the observations. In Fig.~\ref{fig:hh46_grid}, $G_{\rm dust}$ for the best-fit model of BB1.5 are plotted.
Fig.~\ref{fig:hh46_result1} and Fig.~\ref{fig:hh46_result2} shows the horizontal distribution of gas temperature ($T_{\rm gas}$) and CO abundance ($X(\rm{CO})$) for the models of V12, B1.0, B1.5, and the Draine field.   
 
 In the best-fit PDR models, the majority of mid-$J$ CO emission is radiated from the surface ($\Delta R \le \sim 10$~AU or $0.1 \leq A_{\rm V} \leq 1$) of the inner dense UV heated cavity walls with 6~$\leq$ log~$n (\mathrm{cm}^{-3}) \leq$~8, $X(\rm{CO})>$ 10$^{-5}$, and $T_{\rm gas}>$  100~K. The CO~$J = 24-23$ transition traces mostly the warm gas ($T_{\rm gas} \geq$~300~K), while CO $J~=~14-13$ transition arises from  both the warm and cool ($T_{\rm gas} \simeq$~100~K) gas. Therefore, the contribution of the cool gas to the flux of CO~$J~=~14-13$ determines the synthesized rotational temperature.

Our V12 model results in a rotational temperature and FIR mid-$J$ fluxes similar to \citet{Visser12} with 30~$\%$ enhanced UV luminosity. Though FUV radiative transfer and chemistry (especially H$_2$ formation rate) of our model are slightly different from those of \citet{Visser12}, synthesized CO fluxes are similar in two models. Our self-consistent PDR model with BB1.0 also shows a rotational temperature similar to that of the V12 model, but seven times larger UV luminosity is required to match the observation. The fitted UV luminosity for BB1.0 (1.0 $L_{\rm UV}^Y$) is same as the value derived from the observational relation of the classical T-Tauri stars (see above). This result indicates that the approximation of gas temperature and the inconsistency of UV field in the gas energetics and chemistry adopted by \citet{Visser12} might underestimate the UV luminosity of the source.

Unlike the V12 model, our self-consistent PDR models with BB1.5 and the Draine field can reproduce the observed fluxes in the mid-$J$ CO transitions ($E_{\rm up} \le 1,800$~K) without additional heating by a shock, which was adopted by \citet{Visser12}, if the UV luminosity is 3.5~$L_{\rm  UV}^{\rm Y}$ (2.4 L$_{\odot}$). Of course, the line fluxes for $J$ levels with $E_{\rm up} >$~1,800~K cannot be reproduced by the PDR, indicative of shock contribution in the high-$J$ CO lines. However, the important point here is that a self-consistent calculation of PDR could be important to constrain the UV radiation field associated with the accretion process in an embedded protostar.

Our PDR model with BB1.0 has a lower gas temperature than that of V12 model for the same UV luminosity. A higher UV luminosity increases the gas temperature, but it also reduces the CO abundance near the surface. Hence, the model with BB1.0 needs about seven times larger UV luminosity to produce similar fluxes to V12 model. 

BB1.5 has two times lower photodissociation rate of CO than the Draine field. The best fit model with BB1.5    has a slightly lower gas temperature (by about 10\%) but a slightly higher CO abundance  than the best fit model with the Draine field (see  Fig. \ref{fig:hh46_result2}), which results in similar CO fluxes.

 Generally, in a dense PDR (log $n \geq$~6), a higher $G_0/n$ results in a higher gas temperature and a lower CO abundance near the surface. Because the density profile and FUV strength  both follow the inverse square law of the radius, $G_0/n$ is almost constant near the surface, and $\left<A_{\rm V}\right>$ decreases as z is lowered. Therefore, the CO abundance near the surface decreases toward lower z.  However, along the surface of the outflow cavity walls, the CO abundance sharply increases from $n \sim 10^6$~cm$^{-3}$ (z = 500 AU; the green line in Fig.~\ref{fig:hh46_result2}) inward to reach X(CO) $\gtrsim 10^{-5}$, where the FIR mid-$J$ CO emission is radiated. 

\begin{figure*}
\includegraphics[width=0.9 \textwidth]{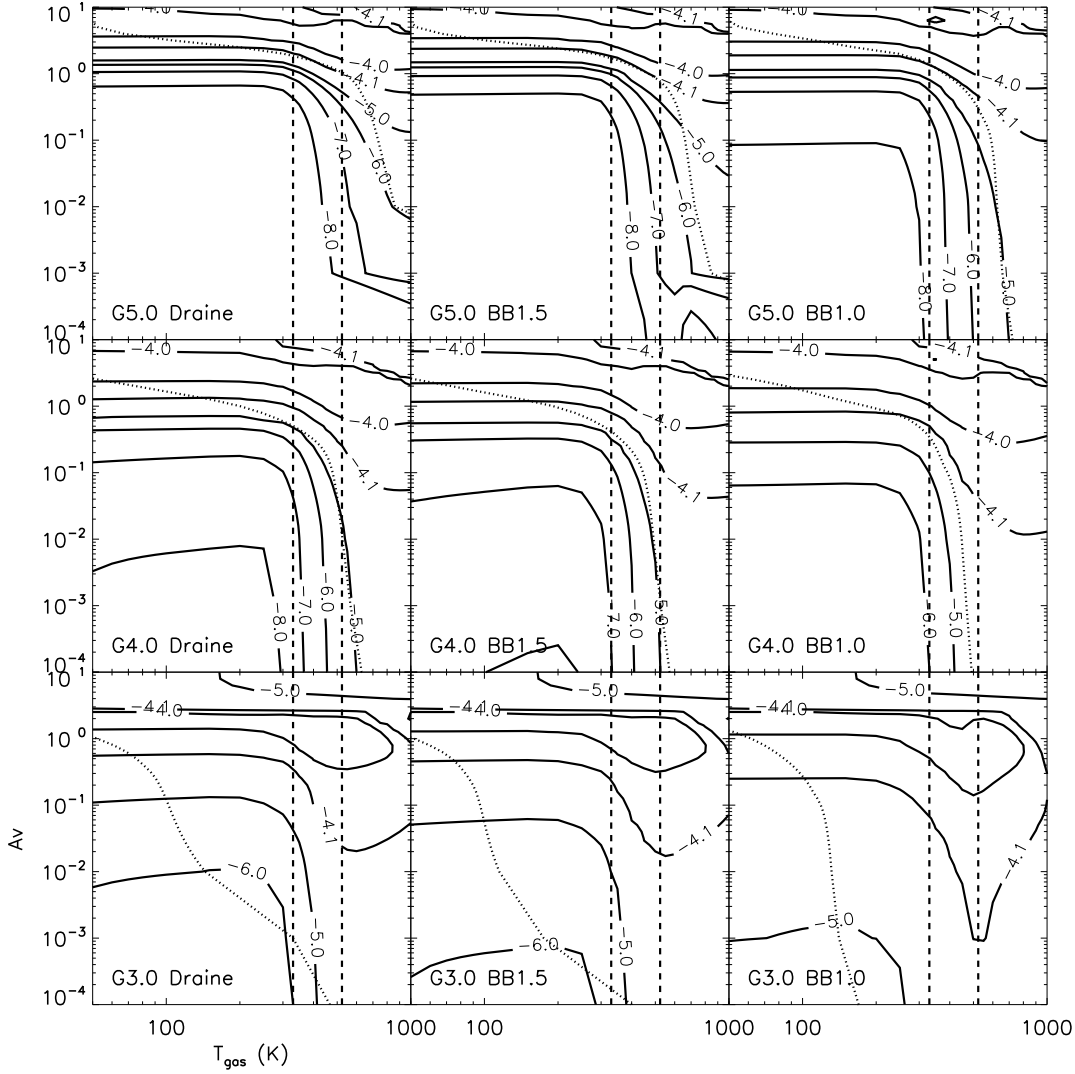}
\caption {Distribution of CO abundance in the domain of $A_{\rm V}$ and $T_{\rm gas}$ for a given  density (log~$n$~=~7~cm$^{-3}$) and $G_0$. The FUV strength (in a log scale) and the type of UV radiation field are presented inside boxes. Contour lines indicate the CO abundance respect to the total hydrogen number density in logarithmic scale. Dotted curves represent the gas temperature of 1D models in Sec. \ref{sec:co1d}, and two vertical lines indicate the gas temperature reproduce the rotational temperature of 300~K for log~$n$~=~7 (330~K) and log~$n$~=~6 (523~K).}\label{fig:co_chem}
 \end{figure*}

Distributions of CO abundance in the domain of $A_{\rm V}$ and $T_{\rm gas}$ for a given FUV strength and gas density (log $n$ = 7 cm$^{-3}$) are plotted in Fig. \ref{fig:co_chem}. For log~$G_0/n \sim -3$ (G4.0; middle row), near the surface (low $A_{\rm V}$), there is the abundance jump around  the gas temperature of a few hundred~K. For example, the model of BB1.5 with the FUV strength of 10$^4$ ISRF (G4.0 BB1.5) has an abundance below 10$^{-7}$ at $T_{\rm gas} <$~300~K, but has the abundance above 10$^{-5}$ in the gas temperature higher than 500~K. In this temperature region, CO forms fast through following reactions \citep{Burton1990}:
\begin{eqnarray}
\mathrm{O} + \mathrm{H_2} &\longrightarrow & \mathrm{OH}  + \mathrm{H} \\
\mathrm{OH} + \mathrm{C^+} &\longrightarrow & \mathrm{CO^+} + \mathrm{H}\\
\mathrm{CO^+} + \mathrm{H_2} &\longrightarrow & \mathrm{HCO^+} + \mathrm{H}\label{eq:co_form1} \\
\mathrm{HCO^+} + \mathrm{e^-}  &\longrightarrow & \mathrm{CO} + \mathrm{H}\label{eq:co_form2}, 
\end{eqnarray}
and near the surface (or higher $G_0/n$), instead of Eq. \ref{eq:co_form1} and \ref{eq:co_form2}, through the reaction,
\begin{equation}
\mathrm{CO^+} + \mathrm{H} \longrightarrow  \mathrm{CO} + \mathrm{H^+}.
\end{equation}

This jump in the CO abundance depends on $G_0/n$ and the radiation field. In a higher $G_0/n$ and the UV radiation field of a higher effective temperature blackbody, the CO abundance jump occurs at a deeper region with a higher gas temperature due to the more efficient photodissociation at the same $A_{\rm V}$. For our best fit model with BB1.5, most fluxes of the mid-$J$ CO lines are emitted from the condition of log~$G_0/n \sim -3$ and $0.1 \leq A_{\rm V} \leq 1.0$, where  the CO abundances increases from $\sim$300~K. 

\section{Summary}\label{sec:summary}
We have developed a self-consistent PDR model with an optimized coordinate system to the embedded protostars with outflow cavities, which reduces a number of grid and a calculation time with no loss of information.  The benchmark test shows that our model agrees with other models in R07. Simple 1D test with our PDR model shows that FIR mid-$J$ CO lines can be emitted from the near the surface (0.1~$\le A_{\rm  V} \le$~1) of dense gas (6 $\le \mathrm{log}\, n\, (\mathrm{cm}^{-3})\le$ 8) exposed to  a high FUV strength (log~$G_0 \ge 3.5$). For the same high density model, a high FUV strength moves the mid-$J$ CO emitting position to the deeper region to reproduce a similar rotational temperature.
 We apply our PDR model to the embedded protostar HH46; our PDR model can provide a high spatial resolution with a small number of grids along the UV heated outflow wall structure. In the application to HH46, we have found that the spectrum of UV radiation field affects the rotational temperature derived from the CO ladder transitions. If we adopt the radiation field of the blackbody of $T_{\rm eff}~=~1.5~\times~10^4$~K or the Draine field with the UV luminosity of 2.4~$L_{\odot}$, we could reproduce the observed fluxes of the rotational transitions with 550~K $< E_{\rm up} < 1,800$~K even without considering a shock contribution. In dense outflow cavity walls ($\mathrm{log}\, n \, \geq 6$) with log~$G_0/n \sim -3$, a higher UV luminosity leads to a higher gas temperature, where the CO abundance increases sharply, resulting in the universal rotational temperature of $\sim$300~K.

\acknowledgments
 We thank N. Evans for commenting on drafts of this manuscript. We are also very grateful to the anonymous referee for helpful comments, which led to improvements in the paper.
This research was supported by the Basic Science Research Program through the National Research Foundation of Korea (NRF) funded by the Ministry of Education of the Korean government (grant No. NRF-2012R1A1A2044689).

\bibliographystyle{aa}
\bibliography{biblio}
\clearpage

\begin{deluxetable}{lrr}
\tablewidth{0pt}
\tablecaption{Binding energies and photo-desorption yields.\label{tab:BePy_grain}}
\tablehead{
\colhead{Species} & \colhead{Binding energy} & \colhead{photo-desorption yield} \\
\colhead{}        & \colhead{$E_{\rm  b}$ (K)} & \colhead{$Y_{\rm i}$ (per UV photon)} 
}
\startdata
CO(gr)      &    855 \tablenotemark{a}   & 2.70 $\times$ 10$^{-3}$ \tablenotemark{b}\\
CO$_2$(gr)  &   2860 \tablenotemark{a}   & 1.00 $\times$ 10$^{-3}$ \tablenotemark{d}\\
H$_2$O(gr)  &   4820 \tablenotemark{a}   & 1.36 $\times$ 10$^{-3}$ \tablenotemark{c}\\
CH$_4$(gr)  &   1360 \tablenotemark{a}   & 1.00 $\times$ 10$^{-3}$ \tablenotemark{d}\\
NH$_3$(gr)  &    880 \tablenotemark{a}   & 1.00 $\times$ 10$^{-3}$ \tablenotemark{d}\\
\enddata
\tablenotetext{a}{\citet{willacy2007}}
\tablenotetext{b}{\citet{oberg2007}}
\tablenotetext{c}{\citet{oberg2009-H2O}}
\tablenotetext{d}{assumed values}
\end{deluxetable}

\begin{deluxetable}{lrlr}
\tablewidth{0pt}
\tablecaption{Initial Abundances\label{initabun}}
\tablehead{
\colhead{Species} & \colhead{Abundance}\tablenotemark{a} & \colhead{Species} & \colhead{Abundance}
}
\startdata
H$_2$ & 5.00E-1 	& CO  & 1.00E-4 \\
He & 1.40E-1 	& N$_2$ & 1.00E-6 \\
N & 2.25E-5 	& C & 7.00E-7 \\
CN & 6.00E-8 	& NH$_3$ & 8.00E-8 \\
H$_3\,^+$ & 1.00E-8 	& HCN & 2.00E-8 \\
S$^+$ & 1.60E-6 		& C$^+$ & 1.00E-8 \\
Si$^+$ & 1.60E-9 		& HCO$^+$ & 9.00E-9 \\
Mg$^+$ & 3.00E-8 		& H$_2$CO & 8.00E-9 \\
Fe$^+$ & 2.00E-8 		& C$_2$H & 8.00E-9 \\ 

H$_2$O(gr) & 2.50E-4	& CS & 2.00E-9 \\
GRAIN & 6.00E-12		&              \\
\enddata
\tablenotetext{a}{Abundance = $\frac{n_{\rm  X}}{n( = ~n_{\rm  H}~+~2~n_{\rm  H_2})}$, where $n_{\rm  X}$ is the number density of species X.
           }
\end{deluxetable}

\clearpage

\end{document}